\documentclass[a4paper,fleqn]{cas-sc}

\newif\ifblind
\blindtrue

\usepackage[authoryear,longnamesfirst]{natbib}
\usepackage{setspace}

\begin{document}
\let\WriteBookmarks\relax
\def\floatpagepagefraction{1}
\def\textpagefraction{.001}

\shorttitle{Does an LLM need the citation sentence?}
\shortauthors{D. A. Bezerra and D. R. Amancio}

\title[mode=title]{Does an LLM Need the Citation Sentence? A Field-Ablation Study of Zero- and Few-Shot Citation Intent Classification}

\author[1]{Davi Alves Bezerra}
\credit{Conceptualization, Methodology, Software, Formal analysis, Investigation, Writing -- original draft, Writing -- review \& editing}

\author[1]{Diego Raphael Amancio}
\cormark[1]
\ead{diego@icmc.usp.br}
\credit{Conceptualization, Methodology, Software, Formal analysis, Investigation, Writing -- original draft, Writing -- review \& editing}

\affiliation[1]{organization={Institute of Mathematics and Computer Sciences, University of São Paulo},
            addressline={Av. Trabalhador São-carlense, 400 - Centro},
            city={São Carlos},
            postcode={13566-590},
            state={São Paulo},
            country={Brazil}}

\cortext[1]{Corresponding author}

\makeatletter
\AtEndPreamble{\NAT@longnamesfalse}
\makeatother

\begin{abstract}
Citation intent classification, labeling why a paper cites another, is usually solved by
fine-tuning an encoder on the inline citation sentence. We ask what a zero- or few-shot large
language model (LLM) actually needs to perform this task, and what prompting costs relative to
supervised classification. Eight open-weight LLMs (8B--20B parameters) are evaluated on ACL-ARC (six classes) and SciCite (three classes) under a field-ablation
design: four prompt-content variants (citation sentence only; sentence plus abstracts and author
lists; abstracts only; author lists only), two prompt styles, and zero- to six-shot conditions,
scored on a fixed comparable sample of 92 ACL-ARC and 1,466 SciCite test rows with paired bootstrap
tests (10,000 resamples). Withholding the citation sentence is uniformly damaging: in all eight
models on both datasets the better text-based variant beats the better metadata-only variant, and
author lists alone reach only 0.09--0.35 macro-F1. Appending metadata to the sentence helps
inconsistently and significantly hurts on SciCite at three-shot. A prompt with explicit label
definitions significantly improves three-shot SciCite macro-F1 ($+0.095$ and $+0.028$ for the two
models tested). Reasoning-mode models gain no accuracy and cost 22--653$\times$ the per-request
latency of direct-answer models. Fine-tuned SciBERT and SPECTER classifiers (0.70 and 0.85 macro-F1
versus 0.62 and 0.76 for the best prompt) beat every prompted configuration; frozen-encoder
baselines do not reliably do so. Metadata-only accuracy does not track the cited paper's citation
count (Spearman $|\rho|\le0.04$), giving no evidence of an author-fame memorization shortcut. The
results argue against metadata-only pipelines and reasoning-oriented models for this task, and
cast prompting as a fallback when no labeled data exists.
\end{abstract}

\begin{highlights}
\item Across eight open-weight LLMs and two benchmarks, withholding the citation sentence always loses to keeping it; author lists alone give 0.09--0.35 macro-F1.
\item Appending paper metadata to the citation sentence helps inconsistently and significantly hurts on SciCite, so richer prompts are not a safe default.
\item A prompt with explicit label definitions raises three-shot SciCite macro-F1 by $+0.095$ and $+0.028$ in the two models on which it was tested.
\item Reasoning-mode LLMs bring no accuracy gain yet cost 22--653$\times$ the per-request latency of direct-answer models on the same GPU.
\item Fine-tuned SciBERT/SPECTER beat every prompted configuration; metadata-only accuracy does not track citation counts, ruling out an author-fame shortcut.
\end{highlights}


\maketitle

\onehalfspacing

\section{Introduction}
\label{introduction}

Scientific citations encode far more than simple bibliographic links; they reflect the nuanced relationships through which authors contextualize their work, acknowledge prior art, and extend existing knowledge. Classifying citation intent (i.e., determining why one paper cites another) is therefore fundamental to downstream tasks such as citation-aware summarization, influence estimation, and citation network analysis \citep{Jurgens2018,Cohan2019}. For network analysis in particular, the practical stakes are high: filtering a citation network by intent substantially alters which papers emerge as central \citep{bezerra2026leveraging}. Consequently, errors in intent classification propagate directly into the bibliometric assessments and impact measures built upon them.

The dominant approach fine-tunes a pretrained encoder on the inline
citation sentence. Large language models (LLMs) prompted zero- or few-shot \citep{Brown2020} offer
an attractive alternative that needs no labeled training set, but two practical design choices are
under-studied: (i) whether the model actually needs the citation sentence, given that paper-level
metadata (title, abstract, authors) is often easier to obtain at scale than clean inline citation
context, and (ii) whether a prompt that explicitly defines each label improves over a minimal
prompt, and whether that benefit is stable across datasets and shot counts. This paper reports a
controlled field-ablation and prompt-style study addressing both questions on two established
citation-intent benchmarks, evaluated across eight open-weight LLMs.

The practical stakes of question (i) go beyond convenience. Fine-tuning an encoder on inline
citation sentences requires hundreds to thousands of hand-labeled examples per benchmark, and
hand-labeling scientific citations is slow, expert-dependent work; more broadly, \citet{WangLiu2021}
report that matching human-label performance on a range of NLU/NLG tasks costs 50-96\% less when
the labels come from GPT-3 rather than human annotators, and semi-supervised approaches such as
\citet{Croce2020}'s GAN-BERT exist specifically to squeeze more accuracy out of the labeled examples
a project can afford by also learning from unlabeled ones. If a zero-shot LLM prompt could reach
usable accuracy from paper-level metadata alone -- abstracts and author lists, both far cheaper to
obtain at scale than clean inline citation sentences with correctly resolved cited-work spans -- it
would sidestep the labeled-data bottleneck (and the inline-extraction bottleneck) entirely, rather
than merely reducing it, exactly the case for classifying citation intent when no annotated training
set exists at all. Because the LLMs evaluated here are pretrained on text that plausibly includes a
share of the scientific literature itself, this question has a second layer: a metadata-only prompt
does not only test topical reasoning over an abstract, it can also surface whatever the model
happens to remember about the specific cited paper or its authors from pretraining
(Section~\ref{related-work-memorization}). Distinguishing "the model can infer intent from content"
from "the model recognizes this particular paper or author" matters for how far metadata-only
prompting could be expected to generalize beyond well-known, heavily cited work, and for whether any
such shortcut would be trustworthy even if it were accurate.

This paper makes four contributions. (1) A controlled field-ablation design for LLM-based citation
intent classification, in which four prompt-content variants are scored on an identical row set, so
that a macro-F1 difference is attributable to the field withheld rather than to a shift in which
examples are evaluated. (2) A cross-model generalization test of the resulting ordering over eight
open-weight instruction-tuned and reasoning LLMs served under identical, inference-time generation
control. (3) A cost-accuracy comparison against fine-tuned and frozen SciBERT/SPECTER baselines
trained on the same benchmarks, including an isolated per-request latency benchmark and multi-seed
supervised results. (4) A battery of memorization and structural checks -- citation count,
author-name repetition, document section, self-citation, and temporal distance -- on the
metadata-only variants. Section~\ref{objectives} states the research questions;
Section~\ref{related-work} positions the study against prior work; Sections~\ref{methodology}
and~\ref{results} report the design and the results; Section~\ref{discussion} summarizes the findings
by research question and discusses their limitations; and Section~\ref{implications} sets out how the
study advances existing work and its theoretical and practical implications.

\section{Research objectives and questions}
\label{objectives}

The overall aim of this study is to establish what a zero- or few-shot LLM actually has to be given
in order to classify citation intent, and what that choice costs relative to the supervised approach
it is meant to replace. Rather than optimizing any one system, we hold a single pipeline fixed and
vary its inputs, its prompt style, and the model behind it, so that each measured difference is
attributable to a design decision a practitioner would actually face. Five objectives, each phrased
as a research question (RQ), follow from that aim.

\begin{enumerate}
\item \textbf{RQ1 (field ablation).} \emph{How much of the task depends on the inline citation
sentence, and can paper-level metadata replace it rather than merely supplement it?} We quantify the
contribution of the citation sentence by comparing four prompt-content variants -- citation text
only, citation text plus paper-level metadata, abstracts only, and author lists only -- and by
asking whether metadata can \emph{replace} the citation sentence rather than merely supplement it.
This objective is what motivates the comparable-sample constraint (Section~\ref{methodology}):
scored on different row sets, a variant's macro-F1 would confound the information withheld from the
prompt with a shift in which examples survive a missing-metadata filter, and metadata coverage is
precisely the confound most likely to distort this particular comparison.

\item \textbf{RQ2 (prompt style).} \emph{Does a prompt that defines every label outperform a minimal
one, and consistently enough to be adopted as a default?} We compare a minimal
task instruction (\emph{baseline}) against one that supplies a natural-language definition and an
illustrative sentence for every label (\emph{defined}), holding labels, demonstrations, and evaluated
rows fixed, and ask not only whether the richer prompt wins but whether it wins consistently enough
across datasets and shot counts to be adopted as a default (Section~\ref{results}).

\item \textbf{RQ3 (cross-model generalization).} \emph{Does the field-ablation ordering established on
one model hold across instruction-tuned and reasoning LLMs of different sizes?} A field-ablation
ordering established on one model is a claim about that model until it is checked on others, so we
evaluate a roster chosen to vary reasoning behavior and parameter count independently rather than to
maximize leaderboard standing, and re-test the ordering across all eight models
(Section~\ref{results-crossmodel}). Serving every model locally under identical generation-control
mechanisms (Section~\ref{methodology-inference}) is part of this objective: it keeps a between-model
difference from being an artifact of a provider's undocumented and silently drifting sampling
defaults.

\item \textbf{RQ4 (prompting versus supervised classification).} \emph{When labeled data exists, how
does zero/few-shot prompting compare with supervised encoders on accuracy \emph{and} inference cost?}
We train
SciBERT and SPECTER classifiers on each benchmark's own training split in two regimes -- frozen
embeddings and light fine-tuning -- score them on the same rows as the prompted models
(Section~\ref{results-baselines}), and pair that accuracy comparison with an isolated per-request
latency benchmark (Section~\ref{results-cost}). The intent is to answer a deployment question --
which approach should be used, given what labeled data and what compute budget exist -- rather than
a leaderboard one.

\item \textbf{RQ5 (shortcut audit).} \emph{Is whatever accuracy the metadata-only variants reach
explained by memorized bibliographic knowledge or citation-network regularities rather than by the
prompt content?} Any accuracy the
metadata-only variants do reach could come from inference over the supplied content, from memorized
bibliographic knowledge about the specific paper or its authors, or from citation-network regularities
that have nothing to do with the prompt's content. We therefore test V4 accuracy against the cited
paper's own citation count, against author-name repetition internal to each corpus, and against
document section, self-citation, citing-cited temporal distance, and repeated citing-cited pairs
(Sections~\ref{discussion-memorization} and~\ref{discussion-structural}). The distinction matters
practically as well as scientifically: a metadata-only pipeline resting on either shortcut would not
generalize past well-known work, and would inherit the authorship-bias risk documented in
LLM-assisted peer review \citep{MacharlaVasu2026,Howell2025,Sajeva2026}.
\end{enumerate}

Three things are deliberately outside these objectives. We do not pursue state-of-the-art accuracy on
either benchmark: the supervised baselines are competent and standard rather than competitive with
the best published systems (Section~\ref{results-baselines}), because the study varies the inputs to
a fixed pipeline instead of competing on a leaderboard. We do not re-engineer the instruction wording
per field variant, so that a variant's score reflects the informational content of the fields it
supplies rather than how much prompt tuning each variant received (Section~\ref{methodology}). And we
neither fine-tune the LLMs themselves nor evaluate closed, API-served models, both to keep generation
control in our own hands and because reproducibility on unversioned proprietary systems is exactly
the difficulty \citet{Fogelson2025} document for this task.

\section{Related work}
\label{related-work}

\subsection{Citation function and intent classification}
\label{related-work-cic}

Citation counts treat every citation as equivalent, yet citing behaviour is heterogeneous:
\citet{Bornmann2008}'s review of citing-behaviour studies concludes that citations serve rhetorical,
social and cognitive functions that a raw count conflates. Citation content and context analysis
emerged in information science precisely to recover that heterogeneity by reading the text around a
citation \citep{Zhang2013,HernandezAlvarez2016}, and automatic classification of \emph{why} a paper
cites another dates back to rule- and feature-based systems over rhetorical citation categories
\citep{Teufel2006}. \citet{Kunnath2021}'s meta-analysis of semantic citation classification
catalogues the resulting proliferation of annotation schemes, corpora and feature sets, and
\citet{Zhang2025survey} survey the deep-learning phase of the same literature from a bibliometric
perspective. The two benchmarks used here established the modern supervised setting: ACL-ARC
\citep{Jurgens2018} annotates computational-linguistics citation sentences with a six-way intent
scheme (background, motivation, uses, extension, comparison-or-contrast, future work), while SciCite
\citep{Cohan2019} introduces a coarser three-way scheme (background, method, result) over a larger,
multi-domain corpus and pairs it with \emph{structural scaffolds} -- auxiliary prediction tasks
(citation worthiness, section title) that regularize a shared encoder during fine-tuning.
\citet{Lauscher2022} later showed that realistic citations are often multi-sentence and multi-label,
a setting that the single-sentence, single-label benchmarks used here do not capture.

Both benchmarks are typically solved by fine-tuning a pretrained transformer encoder, most often
SciBERT \citep{Beltagy2019}, on the inline citation sentence, optionally augmented with the
surrounding context window. Subsequent supervised work has improved on that recipe mainly by adding
information or structure around the encoder: multi-task objectives that share an encoder across
related citation tasks \citep{Qi2023,Zhao2026}; cited-paper information supplied alongside the
citation context \citep{Ghosal2024}; lexical, WordNet-semantic and part-of-speech features fused with
the context representation \citep{Xu2025}; one-vs-all ensembles of encoder pairs \citep{Paolini2025};
heterogeneous-graph models over citation contexts \citep{Quan2026,Phan2026}; causal backdoor
adjustment against confounded context features \citep{Song2026}; and LLM-generated labels aggregated
by a Borda rule to train an encoder \citep{Barchane2026}. 
More broadly, the local textual context of a word or span has long been the key signal for resolving its meaning: in word co-occurrence networks, the consistency of a word's neighborhood decreases with its ambiguity~\citep{Amancio2012using}, and this local structure underpins embedding- and network-based word sense induction~\citep{Correa2019} as well as recent LLM-based attempts at the same task~\citep{mosolova2025llm}. Citation intent can be seen as an analogous problem in which the citation sentence plays the role of that local context.
Semi-supervised variants further reduce the amount of labeled data required: GAN-BERT
\citep{Croce2020} adds an adversarial generator/discriminator pair on top of a BERT-style encoder so
that unlabeled citation sentences also contribute gradient signal during fine-tuning, and
\citet{bezerra2026leveraging} apply this approach to citation intent and show that filtering a
citation network by the predicted intent changes which papers rank as central.
Section~\ref{results-baselines} reports fine-tuned and frozen-embedding SciBERT/SPECTER baselines on
the same splits used here, as a supervised reference point for the prompting results.

\subsection{LLMs for citation intent classification}
\label{related-work-llm-cic}

Prompting language models for citation intent, rather than fine-tuning an encoder on it, begins with
CitePrompt \citep{Lahiri2023}, which recasts the task as cloze-style prompt learning over a
pretrained encoder and reports 53.9\% F1 on ACL-ARC zero-shot, rising to 63.6\% and 67.0\% at five-
and ten-shot -- establishing that the task is partially solvable without task-specific gradient
updates. \citet{Kunnath2023} systematically compare prompting strategies on ACL-ARC and ACT2 --
promptless fine-tuning, fixed-prompt and dynamic-context prompt tuning, and tuning-free zero-shot
GPT-3.5 -- and find that the strategies which update encoder parameters win, while zero-shot GPT-3.5
performs well on ACT2 but poorly on ACL-ARC. From the information-science side,
\citet{Nishikawa2024} compare GPT annotation of citation contexts against human annotators and
conclude that the LLM matches or exceeds human consistency but not human predictive accuracy, so that
it cannot yet replace annotators outright.

Generative LLM prompting on the two benchmarks used here is more recent and still unsettled.
\citet{Koloveas2025} benchmark twelve model variants across five open LLM families (Llama 3, Mistral,
Phi-3, Gemma 2, Qwen2/2.5) on ACL-ARC and SciCite under zero-, one-, few-, and many-shot in-context
learning, then fine-tune the best-performing model on the same data: the best in-context
configuration (Qwen2.5-14B) reaches 63.7\% macro-F1 on ACL-ARC and 78.3\% on SciCite, and fine-tuning
the same model raises these to 68.5\% and 86.8\%, echoing our own finding
(Section~\ref{results-baselines}) that a trained classifier ultimately beats prompting.
\citet{Vodicka2026} instead vary the citation-context \emph{window} fed to five smaller open models
(Mistral 7B, Orca 2-7B, Llama 3.1-8B, Falcon 7B, plus a fine-tuned SciBERT baseline) -- the full
paragraph, a fixed word window, or one-to-three surrounding sentences -- and report zero-shot
macro-F1 ranging from 1.4\% to 47.6\% depending on model and window, with a fine-tuned Falcon 7B
(73.3\%) again the strongest configuration overall. \citet{Fogelson2025} take a
reproducibility-focused view across several proprietary and open systems and report substantial
disagreement between state-of-the-art LLMs on the same citation-intent inputs, attributing part of
the gap to opaque, unversioned proprietary models -- a concern our exclusively open-weight, locally
served setup (Section~\ref{methodology}) is designed to avoid. Two 2026 studies in
\emph{Scientometrics} extend LLM use beyond single-label classification: CiteFuncRanker
\citep{Wang2026} reformulates citation function analysis as pairwise ranking over candidate functions
so that an LLM can express several functions per citation, and \citet{Nguyen2026} use LLM prompting
to extract and semantically enrich the citation context itself before it is analysed.

\paragraph{Positioning of this study.} Table~\ref{tbl:positioning} summarizes how the present study
relates to this line of work, and three things distinguish it. First, none of the prior LLM studies
varies \emph{what information the model is given} in place of the citation sentence: CitePrompt,
\citet{Kunnath2023} and \citet{Koloveas2025} fix the citation context and vary the prompting
strategy, model or shot count, and \citet{Vodicka2026} vary only how wide a window around the
citation is included. Paper-level metadata does appear in the \emph{supervised} literature
\citep{Ghosal2024,Xu2025,Quan2026}, but always as an \emph{addition} intended to improve a model that
already has the citation sentence, never as a candidate \emph{replacement} for it. Whether an LLM can
classify citation intent from metadata alone, when the citation sentence is withheld entirely, is
the question this paper isolates. Second, we hold the evaluated rows fixed across every variant
(Section~\ref{methodology}), so that a macro-F1 difference is attributable to the field removed
rather than to a shift in which examples survive a missing-metadata filter -- a control that matters
precisely because metadata coverage is the confound most likely to distort such a comparison. Third,
the prior LLM studies report accuracy alone; we pair accuracy with an isolated per-request latency
benchmark across reasoning and direct-answer models and with multi-seed supervised baselines, so
that the deployment question -- prompt or fine-tune, and with which kind of model -- can be answered
on cost as well as accuracy.

\begin{table*}[width=\textwidth,cols=6,pos=t]
\caption{LLM-based citation intent classification studies on the benchmarks used here, and how the
present study differs. ``Same rows'' asks whether every compared condition is scored on an identical
evaluated row set.}
\label{tbl:positioning}
\begin{tabular}{@{}p{0.12\textwidth} p{0.20\textwidth} p{0.13\textwidth} p{0.28\textwidth} p{0.07\textwidth} p{0.05\textwidth}@{}}
\toprule
Study & Models & Benchmarks & Factor varied & Same rows & Cost \\
\midrule
\citet{Lahiri2023}    & prompt-tuned SciBERT           & ACL-ARC, SciCite & shot count                              & n/a & no \\
\citet{Kunnath2023}   & SciBERT prompt tuning; GPT-3.5 & ACL-ARC, ACT2    & prompting strategy                      & n/a & no \\
\citet{Koloveas2025}  & 12 variants, 5 open families   & ACL-ARC, SciCite & model, shot count, fine-tuning          & n/a & no \\
\citet{Vodicka2026}   & 4 open 7--8B LLMs, SciBERT     & ACL-ARC, AC3     & context window, fine-tuning             & not stated & no \\
This study            & 8 open 8--20B LLMs (3 reasoning) & ACL-ARC, SciCite & prompt fields, prompt style, shot count, model & yes & yes \\
\bottomrule
\end{tabular}
\end{table*}

\subsection{In-context learning and demonstration selection}

Zero- and few-shot prompting of LLMs \citep{Brown2020} recasts classification as next-token
prediction conditioned on a natural-language instruction and, optionally, a handful of labeled
demonstrations, without any gradient update. A body of work has shown that in-context learning
(ICL) is considerably more sensitive to how demonstrations are chosen and ordered than the
"few-shot learning" framing suggests: \citet{Zhao2021} show that ICL predictions carry systematic
majority-label and recency biases inherited from the demonstration set, correctable with a
calibration step; \citet{Lu2022} show that the \emph{order} of an otherwise-fixed set of
demonstrations alone can swing accuracy by tens of points on some tasks. The configuration studied
here adds a label-\emph{coverage} instance of this same broader sensitivity: ACL-ARC's six-class label space
leaves half of its classes undemonstrated at three-shot under a naive round-robin selection, so
which examples end up in the prompt -- not just how many -- is a free design variable whenever the
number of classes approaches the shot count. We therefore control it explicitly and report a
full-coverage six-shot condition alongside the three-shot one (Section~\ref{methodology}), although
our ACL-ARC evaluation sample proves too small to resolve whether the coverage gap costs accuracy:
the three-shot-versus-six-shot difference does not survive a paired bootstrap
(Section~\ref{results}). Separately, chain-of-thought prompting
\citep{Wei2022,Kojima2022} showed that inducing an intermediate reasoning trace before the final
answer improves LLM accuracy on arithmetic and commonsense tasks; the reasoning-oriented models
evaluated here (Section~\ref{results-crossmodel}) emit such a trace by default, and we find it
costs more than it earns on this classification task once the model's output-token budget is
fixed.

\subsection{Open-weight LLMs evaluated}

The instruction-tuned and reasoning models compared in this study span several open-weight model
families released in 2024--2025: Meta's Llama 3 \citep{Llama3_2024}, Google's Gemma 3
\citep{Gemma3_2025}, Alibaba's Qwen3 \citep{Qwen3_2025}, whose chat template can switch a
reasoning phase on or off per request, OpenAI's gpt-oss \citep{GPTOSS_2025}, Microsoft's
Phi-4-reasoning \citep{Phi4Reasoning_2025}, and DeepSeek's R1-distilled reasoning models
\citep{DeepSeekR1_2025}, which transfer long chain-of-thought reasoning traces from a larger teacher
model into smaller open-weight checkpoints via supervised distillation. All models are
served locally rather than through a vendor API, which is what makes the token-budget-vs-accuracy
trade-off in Section~\ref{results-crossmodel} directly measurable and controllable.

\subsection{What do LLMs know about papers and authors?}
\label{related-work-memorization}

Variants V3 and V4 (Section~\ref{methodology}) ask whether an LLM can infer citation intent from an
abstract or an author list alone, with no citation sentence at all. Because LLMs are pretrained on
web-scale text that plausibly includes a substantial fraction of the scientific literature itself,
this is not simply a test of topical reasoning: it is also a test of what the model \emph{remembers}
about specific papers and authors from pretraining. \citet{Niimi2026} prompts GPT-4.1 to reproduce
full bibliographic records (title, authors, venue, year) for 100 computer-science papers and finds a
strong, log-linear relationship between a paper's citation count and the factual accuracy of what
the model recalls about it, with records reproduced almost verbatim beyond roughly 1{,}000
citations; the extended version of that study \citep{Niimi2026ext} locates an inflection point
around 90 citations, below which the model largely confabulates plausible-sounding records rather
than retrieving memorized ones, and a saturation point around 1{,}200 citations, and shows that
titles and first authors are recalled well before co-authors, venue, or page numbers, which require
substantially higher citation counts to surface reliably. \citet{Shen2026} probe 17 LLMs with multiple-choice
questions about 549 computer-science papers' titles, authors, methods, and venues, and find that
models' aggregate recall correlates with the papers' later citation counts (ensemble Spearman
$\rho=0.15$, $p<0.001$), with author-identity questions the single most discriminative probe --
consistent with prominent authors' papers receiving disproportionate exposure during pretraining.
Together, these results predict that an author-only prompt (V4) should carry usable signal only for
the minority of cited papers whose authors are prominent enough, or whose specific paper is cited
often enough, to have been memorized above this threshold -- which is not the typical citation in
ACL-ARC or SciCite, both drawn largely from mid-tier venues rather than the highest-citation tail of
the literature. We return to this prediction against our own V3/V4 results in
Section~\ref{discussion-memorization}.

A second body of work cautions that \emph{when} an LLM does respond to author identity, the response
is not necessarily principled reasoning about the cited work. In LLM-assisted peer review,
\citet{MacharlaVasu2026} and \citet{Howell2025} both find that revealing author or institutional
identity shifts LLM judgments of otherwise-identical manuscript content -- toward higher acceptance
for prestigious affiliations and senior authors, and toward rejection for lower-status ones -- even
though the underlying text is unchanged; \citet{Howell2025} report that revealing author identity
alone shifts simulated-reviewer rejection recommendations by roughly 25\% on identical manuscripts,
and \citet{Sajeva2026} find that four LLMs scoring scientific abstracts change their scores when the
same abstract is attributed to famous authors or to fabricated authoritative ones.
If an author-only citation-intent prompt were to perform well, this literature suggests the
mechanism would need scrutiny before being trusted: a model could just as easily be pattern-matching
on perceived author status as genuinely inferring why the citing paper invokes the author's work.

\section{Methodology}
\label{methodology}

\subsection{Datasets and the comparable-sample constraint}

We use the test splits of ACL-ARC \citep{Jurgens2018} (139 rows, 6 intent classes) and SciCite
\citep{Cohan2019} (1{,}861 rows, 3 intent classes). Cited-paper metadata is incomplete in both
corpora -- notably, only 59.2\% of ACL-ARC test rows have a cited-paper abstract available even
after enrichment through a cascading lookup across bibliographic sources (Semantic Scholar,
OpenAlex, Crossref, PubMed, arXiv), leaving 40.8\% of cited papers without an abstract. Rather
than imputing a placeholder, we restrict the field-ablation experiments to a \emph{comparable
sample}: the subset of rows for which every field used by \emph{any} tested variant is
simultaneously present (92/139 ACL-ARC rows, 66\%; 1{,}466/1{,}861 SciCite rows, 79\%). Every
variant for a given dataset is therefore scored on the identical row set, so that a change in
macro-F1 reflects the removed/added field rather than a change in which examples are being
evaluated.

\subsection{Field-ablation variants}

Four prompt-content variants are compared per dataset: V1 \emph{text\_only} (the inline citation
sentence alone); V2 \emph{text\_meta} (citation sentence plus citing/cited abstracts and author
lists); V3 \emph{abstracts} (citing/cited abstracts only, no citation sentence); and V4
\emph{authors} (author lists only, no citation sentence). Each variant is run at zero-, one-, and
three-shot (\emph{few}), under two prompt styles: \emph{baseline} (task instruction and label set
only) and \emph{defined} (baseline plus an explicit natural-language definition and example for
every label). All other conditions (labels, examples, dataset) are held fixed across styles. The
illustrative sentence attached to each label definition under \emph{defined} is taken from each
benchmark's own published annotation scheme; we verified by exact-phrase search that none of these
sentences occurs anywhere in either evaluated test split (the ACL-ARC examples that do occur in the
corpus appear only in its training split), so the definition prompt carries no test-set leakage.
Appendix~\ref{app:prompts} gives the full prompt template, both system instructions and the label
definitions verbatim; Appendix~\ref{app:variants} lists the exact dataset columns each variant
renders.

The same generic instruction wording is used for every field variant. We deliberately did not
tailor the instruction to say which fields a given variant supplies, so that a variant's score
reflects the informational content of the fields themselves rather than the extent to which we
re-engineered the prompt around them; the cost of this choice is that sparser variants are never
given a prompt optimized for their sparsity, which is a component of, not a confound in, the
ablation being measured.

Throughout, we follow the in-context learning convention of \citet{Brown2020}: a $K$-shot condition
supplies $K$ demonstrations in total, at most one per label, rather than $K$ demonstrations per label
as in the meta-learning and prompt-based fine-tuning literatures. Few-shot demonstrations are
selected by round-robining across the label space (one example per
label per round) from a seed-shuffled copy of the label order, rather than the dataset's declared
order, so that whichever labels get demonstrated at low shot counts is not silently determined by
declaration order. This matters because ACL-ARC has six intent classes: at three-shot, only half
of the classes can be demonstrated at all. We therefore add a fourth shot count for ACL-ARC only
-- six-shot, one demonstration per class -- giving a condition with full label coverage; SciCite
(three classes) already reaches full coverage at three-shot and needs no such addition.
Appendix~\ref{app:shots} reports the demonstrated-label set this procedure produces at each shot
count.

\subsection{Models and inference}
\label{methodology-inference}

\begin{sloppypar}
Eight instruction-tuned and reasoning LLMs were evaluated: \texttt{gpt-oss-20b}
\citep{GPTOSS_2025}, \texttt{Llama-3.1-8B-Instruct} \citep{Llama3_2024}, \texttt{gemma-3-12b-it}
\citep{Gemma3_2025}, \texttt{Qwen3-8B} and \texttt{Qwen3-14B} \citep{Qwen3_2025},
\texttt{Phi-4-reasoning} \citep{Phi4Reasoning_2025},
\texttt{DeepSeek-R1-\allowbreak Distill-\allowbreak Llama-8B} and
\texttt{DeepSeek-R1-\allowbreak Distill-\allowbreak Qwen-14B} \citep{DeepSeekR1_2025}. Models were
served locally with vLLM \citep{Kwon2023}, one at a time per GPU, with an output-token budget
tuned per model family (32 tokens for direct-answer models; 512 tokens for direct-decoding
reasoning models; a 1536-token reasoning budget under \emph{bounded\_then\_force}, see below). Only
\texttt{gpt-oss-20b} was evaluated on the full 16-configuration sweep (both prompt styles $\times$
all four variants $\times$ both datasets); for the remaining seven models, \emph{defined} was used
throughout, with \emph{baseline} additionally re-run on SciCite at three-shot only to confirm
\emph{defined}'s advantage generalizes beyond \texttt{gpt-oss-20b} (Section~\ref{results}).
\end{sloppypar}

The roster was chosen to vary two properties independently, rather than to chase whichever
open-weight models score highest on general-purpose leaderboards. First, whether the model exposes
an explicit reasoning phase before answering: \texttt{Phi-4-reasoning} and both
\texttt{DeepSeek-R1-Distill} variants always reason; \texttt{Qwen3-8B} and \texttt{Qwen3-14B} can
toggle reasoning per request; \texttt{gpt-oss-20b}, \texttt{Llama-3.1-8B-Instruct}, and
\texttt{gemma-3-12b-it} answer directly with no separate reasoning phase. This axis speaks directly
to the field-ablation question itself -- whether spending more inference-time computation on
chain-of-thought helps a model extract citation intent from sparser prompt content, or mainly adds
cost. Second, parameter count: the roster spans 8B (\texttt{Llama-3.1-8B-Instruct},
\texttt{DeepSeek-R1-Distill-Llama-8B}, \texttt{Qwen3-8B}) to 20B (\texttt{gpt-oss-20b}), all
servable locally on a single 32GB GPU, so that differences in accuracy cannot be attributed to
sheer scale alone. All eight are open-weight; every model was served locally rather than through a
provider API, both to keep the generation-control mechanisms below (guided decoding, reasoning
budgets) under our own control and to avoid an API provider's undocumented sampling defaults
silently drifting across runs.

A reasoning model's chain-of-thought can run to hundreds of tokens before it states a label, and a
free-form final answer is not guaranteed to reproduce one of the six ACL-ARC or three SciCite label
strings verbatim. Rather than lean on a post-hoc parser to recover a label from such text, we
constrain generation itself at inference time, through one of three mechanisms selected per model
family according to what its serving stack supports. Under \emph{direct} decoding, the response is
produced in a single turn under grammar-constrained (guided) decoding \citep{Willard2023}, which
restricts every generated token to a prefix of one of the valid label strings; this suffices for models with no
separate reasoning phase to control. Under \emph{disable\_thinking}, a model's own chat-template
toggle for suppressing its reasoning phase is combined with that same single-turn guided decoding.
Under \emph{bounded\_then\_force}, used for reasoning models that expose no such toggle, generation
proceeds in two turns: an initial, unconstrained turn -- capped at a fixed token budget
(\emph{budget forcing}, \citealp{Muennighoff2025}) -- lets the model reason freely up to that limit,
after which a closing \texttt{\textless/think\textgreater} tag is appended if the model has not already emitted one; a
second, grammar-constrained turn then continues generation directly from that closed reasoning,
immediately constrained to a valid label. Under all three mechanisms, a well-formed request is
guaranteed to terminate in exactly one label from the allowed set, so no downstream label-extraction
step is needed.

Two models could not be guaranteed this way. \texttt{gpt-oss-20b}'s serving stack silently drops
structured-output constraints under its tool-use routing irrespective of how the request is framed,
so a guided-decoding request returns unconstrained free text with no error to signal the failure.
\texttt{DeepSeek-R1-\allowbreak Distill-\allowbreak Llama-8B}'s grammar-constrained continuation
intermittently rejected an otherwise well-formed continuation on roughly one in five requests, a
fault that reproduced even at minimum concurrency and was therefore not a scheduling race, and that
we could not resolve within this study. For both, we fall back to unconstrained generation --
respectively a single direct turn and the free first-turn reasoning under
\emph{bounded\_then\_force} -- and recover the label from the resulting free text with a heuristic
parser that scans the response for the last unhedged mention of a label string. For the remaining
four models, generation is guaranteed by construction to terminate in a valid label, and no parsing
is applied. Per-model generation settings, together with the hardware and vLLM configuration used to
serve them, are given in Appendix~\ref{app:serving}.

The primary metric is macro-averaged F1.

\subsection{Statistical testing}
\label{methodology-stats}

Given ACL-ARC's small comparable sample (92 rows), point-estimate differences between two
configurations can be within sampling noise. For every comparison we report as significant or not,
we run a paired bootstrap \citep{Efron1993} over the shared row set: 10,000 resamples (with replacement) of the row
indices, macro-F1 recomputed on each resample for both configurations under the same resampled
indices, yielding a distribution of paired differences from which we report the mean, a
percentile-based 95\% confidence interval, and $P(\Delta \le 0)$. A 95\% CI that excludes zero is
reported as significant at $\alpha=0.05$.

\subsection{Supervised baselines}
\label{methodology-baselines}

As a supervised reference point, we additionally report two encoder classifiers trained on each
benchmark's own training split: SciBERT \citep{Beltagy2019}, a BERT model pretrained on scientific
text, and SPECTER \citep{Cohan2020}, a SciBERT-initialized encoder further trained with a
citation-informed objective. Each is run at two training regimes, giving four supervised baselines
in total: (i) \emph{frozen embeddings}, where the
encoder is used only to produce a fixed sentence embedding for the inline citation text, on top of
which a small multi-layer perceptron head is trained; and (ii) \emph{fine-tuned}, where the last
two transformer layers and pooler are unfrozen and trained jointly with the classification head.
Both regimes are supervised (trained on ACL-ARC's/SciCite's own training split) and therefore not
directly comparable to the zero/few-shot LLM results in terms of what data each approach consumes;
Section~\ref{results-cost} discusses the resulting cost-accuracy trade-off explicitly.

Two controls make the resulting comparison meaningful. Both baselines are evaluated on the same
comparable sample as the LLM runs, not on the full test split, so that a difference between a
supervised and a prompted score cannot come from the two being scored on different rows. And because
ACL-ARC's evaluation sample is small enough for training-run variance to rival the effects being
measured, each ACL-ARC supervised configuration is trained five times under different seeds and
reported as a mean with its observed range; SciCite, an order of magnitude larger, is reported from a
single run after verifying that independent reruns agreed to within 0.03 macro-F1. Appendix
\ref{app:supervised} gives the full hyperparameters for both regimes.

In summary, three groups of systems appear in Section~\ref{results}: (a) the eight prompted
open-weight LLMs of Section~\ref{methodology-inference}, which are the subject of the study; (b) the
four supervised encoder baselines trained here -- frozen and fine-tuned SciBERT and SPECTER -- scored
on the same rows as the LLMs; and (c) for scale only, published results of prior systems on the full
test splits \citep{Cohan2019,Lahiri2023,Koloveas2025,Paolini2025,Vodicka2026}, which were not
re-run here and are therefore not on the same row set. Table~\ref{tbl3} keeps the three groups in
separate blocks.

\section{Results}
\label{results}

\subsection{Does the model need the citation text?}

Table~\ref{tbl1} reports \texttt{gpt-oss-20b} macro-F1 under the \emph{defined}
prompt, on the comparable sample, for every variant/shot/dataset combination (ACL-ARC includes the
additional six-shot, full-label-coverage condition; see Section~\ref{methodology}). V1 (text only)
is the strongest configuration in six of the seven shot/dataset combinations, and effectively tied
for strongest in the seventh (SciCite one-shot, V1 and V2 within 0.000). Author lists alone (V4)
remain close to uninformative (0.09-0.35 macro-F1); abstracts alone (V3) are a meaningfully stronger
metadata-only signal but stay far below the text-based variants throughout. The V1-vs-V2 gap (does
appending metadata to the citation text help?) is directionally consistent with V1 ahead at every
ACL-ARC shot count (+0.10 to +0.14 macro-F1), but does not reach significance at any of them under
the paired bootstrap (all four 95\% CIs include zero; e.g.\ three-shot $\Delta=+0.105$, 95\% CI
$[-0.029,+0.237]$). On SciCite the pattern is more mixed: the gap is negligible and non-significant
at zero- and one-shot ($\Delta=-0.012$ and $\Delta=-0.000$), but small and significant at
three-shot ($\Delta=+0.025$, 95\% CI $[+0.003,+0.047]$) -- the one condition, across both datasets,
where appending metadata to the citation text is measurably worse than the citation text alone.
V1's own ACL-ARC performance is close to flat across shot counts (0.617 $\to$ 0.601 $\to$ 0.585
$\to$ 0.600 for zero/one/three/six-shot), and the small three-shot dip relative to six-shot -- the
one condition where only half of ACL-ARC's six classes are ever demonstrated -- is itself well
within noise ($\Delta=-0.018$, 95\% CI $[-0.110,+0.062]$, $P(\Delta\le0)=0.672$); with this study's
sample sizes and inference mechanism, we do not read a reliable label-coverage effect into this
trend.

\begin{table*}[width=\textwidth,cols=6,pos=t]
\caption{\texttt{gpt-oss-20b} macro-F1 by field variant, shot count, and dataset (\emph{defined}
prompt, comparable sample). Six-shot applies to ACL-ARC only (six classes; SciCite already reaches
full label coverage at three-shot).}
\label{tbl1}
\begin{tabular*}{\tblwidth}{@{} LLCCCC@{} }
\toprule
Shot & Dataset & V1 text\_only & V2 text\_meta & V3 abstracts & V4 authors \\
\midrule
zero    & ACL-ARC & \textbf{0.617} & 0.513 & 0.202 & 0.145 \\
zero    & SciCite & 0.736 & \textbf{0.749} & 0.426 & 0.345 \\
one     & ACL-ARC & \textbf{0.601} & 0.457 & 0.219 & 0.171 \\
one     & SciCite & \textbf{0.728} & \textbf{0.728} & 0.416 & 0.254 \\
few(3)  & ACL-ARC & \textbf{0.585} & 0.477 & 0.209 & 0.119 \\
few(3)  & SciCite & \textbf{0.758} & 0.732 & 0.419 & 0.325 \\
few(6)  & ACL-ARC & \textbf{0.600} & 0.507 & 0.258 & 0.187 \\
\bottomrule
\end{tabular*}
\end{table*}

\subsection{Does a richer prompt help?}

\emph{Baseline} was re-run under the inference-time generation-control mechanisms of
Section~\ref{methodology} for exactly one condition -- SciCite, V1 text\_only, three-shot -- but for
the two models (\texttt{gpt-oss-20b} and \texttt{Phi-4-reasoning}) whose full sweep completed early
enough to allow it. Comparing \emph{defined} against \emph{baseline} V1 macro-F1 on that condition
shows a significant win for \emph{defined} in both: $\Delta=+0.095$ for \texttt{gpt-oss-20b} (0.758
vs.\ 0.663, 95\% CI $[+0.074,+0.117]$) and a smaller but still significant $\Delta=+0.028$ for
\texttt{Phi-4-reasoning} (0.657 vs.\ 0.629, 95\% CI $[+0.010,+0.046]$), paired bootstrap
(Section~\ref{methodology-stats}). \emph{Baseline} was not re-run on ACL-ARC, at other SciCite shot
counts, or for the remaining six models, under these mechanisms; comparing those conditions' old
\emph{baseline} results against the new \emph{defined} results would mix pre- and post-fix inference
behavior, so we do not report them here. \emph{Defined} is the default for the cross-model
comparison below on the strength of these two SciCite results together with its performance in the
V1-vs-V2/V3/V4 comparisons above.

\subsection{Cross-model comparison}
\label{results-crossmodel}

Table~\ref{tbl2} reports each model's mean macro-F1 across all four field variants,
under the \emph{defined} prompt (ACL-ARC: four shot counts, 0/1/3/6, 16 configurations per model;
SciCite: three shot counts, 0/1/3, 12 configurations per model -- see Section~\ref{methodology}), for
all eight evaluated models. \texttt{gemma-3-12b-it} and \texttt{gpt-oss-20b} remain the strongest
models overall, essentially tied on both datasets (ACL-ARC 0.372 vs.\ 0.367; SciCite 0.551 vs.\
0.551), with \texttt{Qwen3-14B} close behind on both (0.366, 0.531). Once generation is constrained
by the mechanisms in Section~\ref{methodology} rather than left to a fixed, unconstrained
output-token budget, the reasoning models are no longer uniformly weaker than the direct-answer
ones: \texttt{DeepSeek-R1-Distill-Qwen-14B} in particular improves from being the weakest model
measured under the old, unconstrained pipeline to a mid-table result (0.342, 0.527) once its
\texttt{context\_length\_exceeded} failures -- caused by a 2048-token context cap that no longer
applies, see Section~\ref{methodology} -- are eliminated, and \texttt{Phi-4-reasoning} lands in the
same mid-table band (0.345, 0.500). \texttt{DeepSeek-R1-Distill-Llama-8B} remains the weakest model
on both datasets (0.174, 0.449), but this is not a reasoning-budget artifact: it is the one model
(besides \texttt{gpt-oss-20b}) for which guided decoding could not be enforced reliably, so its
labels are recovered by the heuristic parser described in Section~\ref{methodology} rather than
guaranteed by construction, and that fallback measurably underperforms guided decoding on this task.

\begin{table}[width=\linewidth,cols=5,pos=t]
\caption{Mean macro-F1 per model, \emph{defined} prompt. ACL-ARC: mean over 4 field variants
$\times$ 4 shot counts (0/1/3/6). SciCite: mean over 4 field variants $\times$ 3 shot counts
(0/1/3). ``Reasoning'' marks models that generate a reasoning trace before answering under the
configuration used here (Appendix~\ref{app:serving}).}
\label{tbl2}
\begin{tabular*}{\tblwidth}{@{} LLCCC@{} }
\toprule
Model & Reference & Reasoning & ACL-ARC & SciCite \\
\midrule
gemma-3-12b-it                  & \citet{Gemma3_2025}        & no  & \textbf{0.372} & \textbf{0.551} \\
gpt-oss-20b                     & \citet{GPTOSS_2025}        & yes & 0.367          & \textbf{0.551} \\
Qwen3-14B                       & \citet{Qwen3_2025}         & no  & 0.366          & 0.531 \\
DeepSeek-R1-Distill-Qwen-14B    & \citet{DeepSeekR1_2025}    & yes & 0.342          & 0.527 \\
Phi-4-reasoning                 & \citet{Phi4Reasoning_2025} & yes & 0.345          & 0.500 \\
Qwen3-8B                        & \citet{Qwen3_2025}         & no  & 0.302          & 0.453 \\
Llama-3.1-8B-Instruct           & \citet{Llama3_2024}        & no  & 0.250          & 0.493 \\
DeepSeek-R1-Distill-Llama-8B    & \citet{DeepSeekR1_2025}    & yes & 0.174          & 0.449 \\
\bottomrule
\end{tabular*}
\end{table}

\paragraph{Does the field-ablation ordering from Table~\ref{tbl1} generalize?} Table~\ref{tbl1}'s
V1$>$V2$>$V3$>$V4 ranking is demonstrated in full, with bootstrap significance testing, for
\texttt{gpt-oss-20b} alone. Table~\ref{tbl6} checks whether it holds across all eight models, at
three-shot (the one shot count common to both datasets). The coarse-grained claim generalizes
without exception: on both datasets and all eight models, the better of the two text-based variants
(V1 or V2) outperforms the better of the two metadata-only variants (V3 or V4) -- citation text,
whether alone or padded with metadata, is never beaten by metadata alone. The finer-grained
direction -- does V1 (text alone) beat V2 (text plus metadata), the basis for the abstract's
"appending metadata usually hurts" claim -- is less universal: V1 ranks above V2 in six of eight
models on \emph{each} dataset, but reverses, consistently on both datasets, for \texttt{Qwen3-8B}
(narrowly) and \texttt{DeepSeek-R1-\allowbreak Distill-\allowbreak Llama-8B} (by a wide margin on
ACL-ARC, 0.128 vs.\ 0.225). The latter is also the one model whose labels rely on the heuristic
parser rather than guided decoding (Section~\ref{methodology-inference}), so its reversal may
reflect noisier label extraction under the sparser V1 prompt rather than metadata genuinely helping
that model; \texttt{Qwen3-8B}'s reversal has no such confound and is small enough (V1 vs.\ V2 within
0.01 macro-F1 on both datasets) to be consistent with noise. We therefore scope the
"appending-metadata-hurts" claim to what was directly tested with bootstrap significance
(\texttt{gpt-oss-20b}, Section~\ref{results}) and report it as the common but not universal pattern
across the full roster, rather than a claim established for every model.

\begin{table}[width=\linewidth,cols=3,pos=t]
\caption{Field-variant ranking by macro-F1, three-shot, \emph{defined} prompt, all eight models.
"text$>$meta" reports whether the better of \{V1,V2\} exceeds the better of \{V3,V4\}.}
\label{tbl6}
\begin{tabular*}{\tblwidth}{@{} LCC@{} }
\toprule
Model & ACL-ARC order & SciCite order \\
\midrule
gpt-oss-20b                     & V1$>$V2$>$V3$>$V4 & V1$>$V2$>$V3$>$V4 \\
gemma-3-12b-it                  & V1$>$V2$>$V3$>$V4 & V1$>$V2$>$V3$>$V4 \\
Llama-3.1-8B-Instruct           & V1$>$V2$>$V3$>$V4 & V1$>$V2$>$V3$>$V4 \\
Qwen3-14B                       & V1$>$V2$>$V3$>$V4 & V1$>$V2$>$V3$>$V4 \\
DeepSeek-R1-Distill-Qwen-14B    & V1$>$V2$>$V3$>$V4 & V1$>$V2$>$V3$>$V4 \\
Phi-4-reasoning                 & V1$>$V2$>$V3$>$V4 & V1$>$V2$>$V3$>$V4 \\
Qwen3-8B                        & V2$>$V1$>$V3$>$V4 & V2$>$V1$>$V4$>$V3 \\
DeepSeek-R1-Distill-Llama-8B    & V2$>$V1$>$V4$>$V3 & V2$>$V1$>$V3$>$V4 \\
\bottomrule
\end{tabular*}
\end{table}

\subsection{Error analysis}
\label{results-errors}

\begin{figure}
\centering
\includegraphics[width=0.7\linewidth]{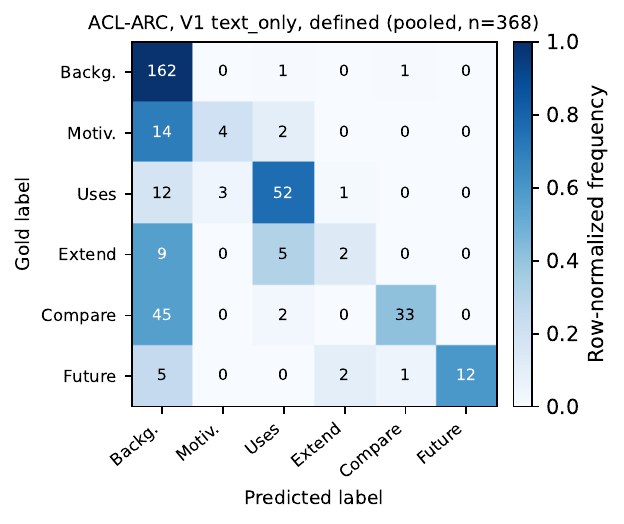}
\caption{Row-normalized confusion matrix, ACL-ARC, \texttt{gpt-oss-20b}, V1 text\_only, defined prompt, pooled across shots.}
\label{fig:cm-acl}
\end{figure}

\begin{figure}
\centering
\includegraphics[width=0.7\linewidth]{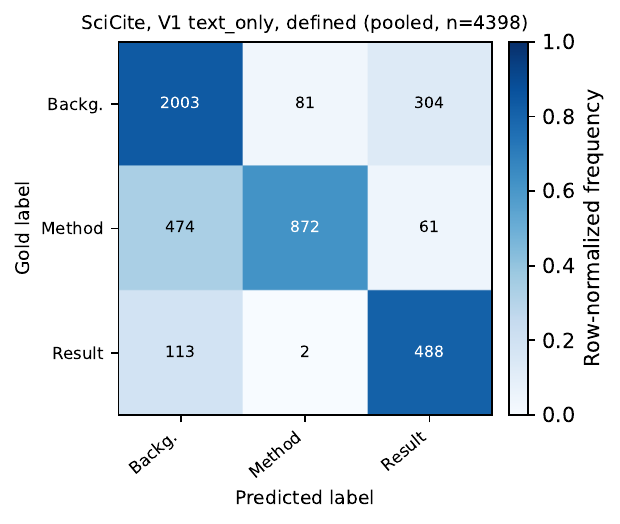}
\caption{Row-normalized confusion matrix, SciCite, \texttt{gpt-oss-20b}, V1 text\_only, defined prompt, pooled across shots.}
\label{fig:cm-scicite}
\end{figure}

Figures~\ref{fig:cm-acl} and~\ref{fig:cm-scicite} show row-normalized confusion matrices for
\texttt{gpt-oss-20b}, V1 text\_only, \emph{defined} prompt, pooled across all shot counts on each
dataset (368 ACL-ARC predictions, 4,398 SciCite predictions; zero invalid predictions in either
pool). On ACL-ARC, \emph{background} is still the class errors most often collapse into -- 43 of
80 \emph{comparison-or-contrast} instances are misclassified as \emph{background} -- but the bias
is markedly less pervasive than a simple "guess background" strategy would produce:
\emph{motivation} errors are now spread across \emph{uses}, \emph{extension}, and
\emph{comparison-or-contrast} rather than concentrating on \emph{background} (5 of 20
\emph{motivation} instances misclassified as \emph{background}), and \emph{background} itself is
predicted correctly in 148 of 164 cases. \emph{Extension} remains the worst-performing class in
absolute terms (5 of 16 correct), consistent with it being one of ACL-ARC's rarest labels (3.7\% of
the full dataset); notably, this is not a demonstration-coverage effect, since \emph{extension} is one
of the three classes the seed-shuffled round-robin does demonstrate at three-shot, alongside
\emph{motivation} and \emph{uses} (Appendix~\ref{app:shots}). On SciCite, the dominant confusion runs
the other way: 427 of 1,407 \emph{method} instances are misclassified as
\emph{background}, and 355 of 2,388 \emph{background}
instances are misclassified as \emph{result} -- a broadly similar, more symmetric error pattern to
ACL-ARC's, consistent with SciCite's larger, more balanced label distribution and the model's
substantially higher overall accuracy there. The remaining concentration of ACL-ARC errors around
\emph{background} is consistent with the field-ablation results above: a model with only the
citation sentence to go on will occasionally default to the majority class on genuinely ambiguous
cases, but no longer to the point of masking distinctions between the other five classes, which
aggregate macro-F1 alone would not reveal.

\subsection{Comparison with supervised baselines}
\label{results-baselines}

Table~\ref{tbl3} adds SciBERT and SPECTER classifiers trained on each dataset's own training split
(Section~\ref{methodology-baselines}) to the best zero/few-shot LLM result from
Table~\ref{tbl1}/\ref{tbl2} (V1 text\_only, macro-F1 maximized over shot count, per dataset). Two
methodological points govern how these rows were produced. First, the supervised classifiers are
scored on the \emph{same comparable sample} as the LLM results rather than on the full test splits,
so the comparison is over an identical row set; scoring them on the full split instead moves ACL-ARC
supervised macro-F1 by up to 0.07 and SciCite by at most 0.011, so this correction matters
substantially on the smaller benchmark. Second, ACL-ARC supervised results proved highly unstable
across runs -- re-training fine-tuned SPECTER under identical hyperparameters produced macro-F1
between 0.639 and 0.757 across five seeds, a spread wider than several of the differences this table
is used to interpret -- so its supervised entries are reported as a mean over five seeds with the
observed range, rather than as single point estimates. SciCite, whose test split is an order of
magnitude larger, is stable: independent reruns agreed to within 0.03.

On this footing, both \emph{fine-tuned} configurations beat the best LLM result on both datasets, and
robustly so: on ACL-ARC the weakest of five fine-tuned SPECTER runs (0.639) still exceeds the LLM's
0.617, and on SciCite the margin is 0.083--0.094. Frozen-embedding baselines are a different story.
On ACL-ARC, frozen SPECTER (0.520) is clearly worse than prompting, while frozen SciBERT's apparent
advantage is not reliable: its five-seed range (0.595--0.709) straddles the LLM's 0.617, so whether
it wins depends on the run. On SciCite, frozen SciBERT (0.834) is clearly ahead, but frozen SPECTER
(0.763) and the best LLM prompt (0.758) are effectively tied. The conclusion that survives is
narrower than a single-run table would suggest: \emph{fine-tuning} an encoder reliably beats
prompting on both benchmarks, whereas merely freezing an encoder and training a small head on it is
not a dependable win over a zero/few-shot prompt on either -- it depends on the backbone, and on
ACL-ARC, on the training run itself.

\begin{table*}[width=\textwidth,cols=3,pos=t]
\caption{Macro-F1 of the prompted LLM (this study) against the supervised baselines trained here and
against published results. The first two blocks are scored on the \emph{same comparable sample}
(92 ACL-ARC rows, 1{,}466 SciCite rows); ACL-ARC supervised entries are the mean over five seeds with
the observed range in brackets, and SciCite entries are single runs (reruns agreed to within 0.03).
The third block reproduces each paper's own reported macro-F1 on the full test split, under that
paper's own protocol; it is not on the same rows and is given for scale only.}
\label{tbl3}
\begin{tabular*}{\tblwidth}{@{} LCC@{} }
\toprule
System & ACL-ARC & SciCite \\
\midrule
\multicolumn{3}{@{}l}{\emph{Zero/few-shot LLM prompting (this study): V1 text\_only, defined prompt, best shot count}} \\
gpt-oss-20b \citep{GPTOSS_2025}                          & 0.617 & 0.758 \\
\midrule
\multicolumn{3}{@{}l}{\emph{Supervised baselines trained here, scored on the same rows}} \\
Frozen SciBERT \citep{Beltagy2019} + MLP head           & 0.668 [0.595--0.709] & 0.834 \\
Frozen SPECTER \citep{Cohan2020} + MLP head             & 0.520 [0.499--0.533] & 0.763 \\
Fine-tuned SciBERT \citep{Beltagy2019}, last 2 layers   & 0.686 [0.646--0.746] & 0.841 \\
Fine-tuned SPECTER \citep{Cohan2020}, last 2 layers     & \textbf{0.701} [0.639--0.757] & \textbf{0.852} \\
\midrule
\multicolumn{3}{@{}l}{\emph{Published results on the full test split (different rows and protocols; not directly comparable)}} \\
Structural scaffolds, SciBERT \citep{Cohan2019}         & 0.679 & 0.840 \\
CitePrompt, zero-shot \citep{Lahiri2023}                & 0.539 & -- \\
CitePrompt, ten-shot \citep{Lahiri2023}                 & 0.670 & -- \\
Best in-context LLM, Qwen2.5-14B \citep{Koloveas2025}   & 0.637 & 0.783 \\
Fine-tuned Qwen2.5-14B \citep{Koloveas2025}             & 0.685 & 0.868 \\
CiteFusion ensemble \citep{Paolini2025}                 & 0.762 & 0.896 \\
Fine-tuned Falcon 7B \citep{Vodicka2026}                & 0.733 & -- \\
\bottomrule
\end{tabular*}
\end{table*}

\paragraph{Where this sits relative to the published state of the art.} The third block of
Table~\ref{tbl3} reproduces published results on the full test splits. The supervised baselines
trained here are deliberately standard -- a frozen or lightly fine-tuned encoder with a small MLP
head -- and are not competitive with the best published systems on either benchmark: the CiteFusion
ensemble \citep{Paolini2025} reports 0.896 macro-F1 on SciCite and 0.762 on ACL-ARC, fine-tuned
Falcon 7B \citep{Vodicka2026} reaches 0.733 on ACL-ARC, and fine-tuned Qwen2.5-14B
\citep{Koloveas2025} reaches 0.868 on SciCite, all above the encoders used here; \citet{Xu2025} and
\citet{Phan2026} report further gains from feature fusion and graph-based modelling respectively.
That gap is deliberate and does not affect this study's conclusions -- the purpose here is to hold
one pipeline fixed and ablate its \emph{inputs}, not to compete on a leaderboard, and the supervised
rows serve only to place the prompting results on a familiar scale. It does, however, bound how the
first two blocks of Table~\ref{tbl3} should be read: prompting is being compared against a competent
standard classifier rather than against the best available system, and the margin against a
state-of-the-art supervised model would be wider than shown. In the other direction, the prompting
results compare favourably with earlier prompt-based work on the same benchmark: CitePrompt reports
0.539 zero-shot on ACL-ARC, below the 0.617 reached here, and \citet{Koloveas2025}'s best in-context
configuration reaches 0.637 on ACL-ARC and 0.783 on SciCite against our 0.617 and 0.758 -- a gap of
two to three points that lies within the seed and row-set variation documented above -- although all
of these figures are computed on different row subsets and the comparison is therefore indicative
rather than exact. The 2025--2026 systems in that block are the most recent published results on
these benchmarks that we are aware of; none was re-run on the comparable sample, because doing so
would require re-training or re-prompting each system under its own protocol, which the released
artifacts do not uniformly permit, and because the contribution of this study is the controlled
ablation of a fixed pipeline's inputs rather than a new leaderboard entry.

\subsection{Computational cost}
\label{results-cost}

The accuracy gap in Table~\ref{tbl3} has to be weighed against what each approach actually costs.
Because the production sweep behind Tables~\ref{tbl1}--\ref{tbl2} ran two models at a time across
two GPUs over several days, with occasional crash recovery and idle-GPU backfills, its raw wall-clock
timestamps are not a clean cost measurement. We instead ran a dedicated, isolated benchmark after the
fact: each of the eight models was served alone on an otherwise-idle 32GB GPU and timed on one fixed
configuration -- \emph{defined} prompt, V1 text\_only, three-shot -- the full 92-row ACL-ARC test set
and a fixed 200-row SciCite subset, with server start-up and model loading excluded from the timed
interval. Table~\ref{tbl4} reports per-request latency.

\begin{table}[width=\linewidth,cols=4,pos=t]
\caption{Isolated per-request wall-clock latency (\emph{defined} prompt, V1 text\_only, three-shot),
seconds/request, measured alone on a dedicated GPU. Relative column is against the fastest model.}
\label{tbl4}
\begin{tabular*}{\tblwidth}{@{} LCCC@{} }
\toprule
Model & ACL-ARC & SciCite & Relative \\
\midrule
Llama-3.1-8B-Instruct         & 0.021  & 0.017  & 1.0$\times$ \\
Qwen3-8B                      & 0.022  & 0.017  & 1.0$\times$ \\
Qwen3-14B                     & 0.036  & 0.027  & 1.6$\times$ \\
Gemma-3-12B                   & 0.043  & 0.029  & 1.8$\times$ \\
gpt-oss-20b                   & 0.519  & 0.350  & 22$\times$ \\
DeepSeek-R1-Distill-Llama-8B  & 2.098  & 1.394  & 90$\times$ \\
DeepSeek-R1-Distill-Qwen-14B  & 3.870  & 3.436  & 198$\times$ \\
Phi-4-reasoning               & 12.535 & 12.055 & 653$\times$ \\
\bottomrule
\end{tabular*}
\end{table}

The result is that reasoning mode, not parameter count, dominates inference cost -- and within the
reasoning tier, the specific model matters far more than a shared mechanism or token budget would
suggest. The four direct-answer models sit in a tight 0.02--0.04 s/request band that spans 8B to
14B parameters (\texttt{Qwen3-8B} and \texttt{Qwen3-14B} both disable their native thinking mode
for this task; \texttt{Llama-3.1-8B-Instruct} and \texttt{gemma-3-12b-it} never had one). Any model
that generates reasoning tokens before answering is at least an order of magnitude slower per
request regardless of whether that reasoning is unconstrained (\texttt{gpt-oss-20b}, up to 512
free-form tokens, 22$\times$) or explicitly budgeted (the DeepSeek-R1-Distill models and
\texttt{Phi-4-reasoning}, an identical 1536-token \emph{bounded\_then\_force} budget as described in
Section~\ref{methodology-inference}). But that shared mechanism and budget produce wildly different
costs: \texttt{Phi-4-reasoning} is $653\times$ slower than the fastest model and more than
\emph{three times slower again} than \texttt{DeepSeek-R1-Distill-Qwen-14B} (198$\times$), a model of
comparable parameter count under the identical 1536-token budget -- parameter count and nominal
budget alone do not explain this gap, and we did not instrument token-level generation length to
isolate whether \texttt{Phi-4-reasoning} is simply exhausting its reasoning budget on nearly every
request while the DeepSeek-R1-Distill models terminate their reasoning earlier on average, or
whether the two models differ in per-token decoding cost at matched output length. Whatever the mechanism, the practical conclusion is unambiguous: reasoning-oriented models are not
a uniformly priced category -- \texttt{Phi-4-reasoning}'s per-request cost is itself over three
times \texttt{DeepSeek-R1-Distill-Qwen-14B}'s, nearly three orders of magnitude past the fastest
direct-answer models -- and is the least viable prompting choice for this task on cost grounds
alone, independent of its accuracy (Table~\ref{tbl2}). By contrast, training the supervised
baselines in Table~\ref{tbl3} involves no per-query generation at all: the frozen embedding is
computed once per row and cached, and the classification head is a few-thousand- to
few-hundred-thousand-parameter multi-layer perceptron trained for at most 100 epochs with early
stopping; fine-tuning additionally backpropagates through the encoder's last two layers, but still
touches each training row only a handful of times rather than generating hundreds of output tokens
per row at inference time. The practical implication is that prompting only makes sense as a
citation-intent classification strategy when no labeled training data is available at all, or when
a direct-answer (non-reasoning) model is used; once training labels exist, a small fine-tuned
encoder is both more accurate and cheaper per row to run at inference time than repeatedly
prompting an LLM, and reasoning-oriented LLMs in particular are a poor fit for this task on both
axes.

\section{Discussion and limitations}
\label{discussion}

\subsection{Summary of findings by research question}
\label{discussion-summary}

\begin{description}
\item[RQ1 (field ablation).] The inline citation sentence is indispensable. Withholding it costs more
accuracy than any prompt style, shot count or model choice recovers: V1 (text only) is the best
variant in six of seven shot/dataset combinations for \texttt{gpt-oss-20b} (Table~\ref{tbl1}), author
lists alone stay at 0.09--0.35 macro-F1, and abstracts alone, while clearly better than author lists,
remain far below any text-based variant. Metadata therefore cannot \emph{replace} the citation
sentence, and supplementing the sentence with it helps only inconsistently -- non-significantly
negative on ACL-ARC at every shot count and significantly negative on SciCite at three-shot
($\Delta=+0.025$ in favour of text only, 95\% CI $[+0.003,+0.047]$).
\item[RQ2 (prompt style).] The definition-based prompt significantly outperforms the minimal one on
the condition where both were run under the same inference mechanism (SciCite, three-shot) for both
models tested ($+0.095$ and $+0.028$ macro-F1). Its advantage on ACL-ARC is directionally consistent
but not significant on the 92-row sample, so we recommend it as a default rather than report it as
an established, general win.
\item[RQ3 (cross-model generalization).] The coarse ordering -- text-based variants above
metadata-only variants -- holds without exception for all eight models on both datasets
(Table~\ref{tbl6}). The finer claim that text alone beats text plus metadata holds for six of eight
models and reverses for two, one of which is the model whose labels rely on a heuristic parser.
\item[RQ4 (prompting versus supervised classification).] Fine-tuned SciBERT and SPECTER beat the best
prompted configuration on both benchmarks (Table~\ref{tbl3}), and do so robustly across seeds on
ACL-ARC; frozen-encoder baselines do not reliably win. On cost, reasoning-mode LLMs run at
22--653$\times$ the per-request latency of direct-answer LLMs (Table~\ref{tbl4}) with no accuracy
advantage (Table~\ref{tbl2}), and a fine-tuned encoder involves no per-query generation at all.
\item[RQ5 (shortcut audit).] Metadata-only accuracy does not track the cited paper's citation count
($\rho=-0.02$ on ACL-ARC, $\rho=0.04$ on SciCite), model predictions agree with an author-repetition
baseline only modestly above chance, and V3 beats V4 throughout -- the opposite of what paper-level
recognition would produce. Document section and SciCite's key-citation flag do predict correctness,
in the direction of surface-cue matching rather than argumentative understanding
(Sections~\ref{discussion-memorization} and~\ref{discussion-structural}).
\end{description}

\subsection{Content reasoning vs.\ memorized metadata}
\label{discussion-memorization}

The field-ablation results in Section~\ref{results} give a first, indirect answer to the
memorization question raised in Sections~\ref{introduction} and~\ref{related-work-memorization}.
If LLMs' performance on metadata-only prompts were driven mainly by the model recognizing specific,
well-known papers or authors, V4 (author lists) should track the field's most prominent names and
V3 (abstracts) should offer little advantage over having no citation sentence at all, since
paper-level recognition rather than topical content would be doing the work. We observe the
opposite pattern: V3 consistently and substantially outperforms V4 across every dataset and shot
count in Table~\ref{tbl1}, and V4 sits close to the weakest results observed anywhere in this study.
Read through \citet{Niimi2026} and \citet{Shen2026}'s memorization-threshold findings
(Section~\ref{related-work-memorization}), this is the outcome those results would predict for
benchmarks like ACL-ARC and SciCite, whose cited works are drawn overwhelmingly from mid-tier NLP
and general-science venues rather than the small, extremely highly cited tail where LLM memorization
of author-paper associations reliably activates: most cited papers in these test sets likely never
crossed the citation threshold needed for an LLM to have anything specific memorized about their
authors, so V4 largely reduces to guessing from author-name surface features rather than from
substantive knowledge of the cited work. This also means the weak positive signal V4 does carry
above the floor cannot, from this study alone, be cleanly attributed to genuine bibliographic
recognition versus a superficial name-based prior of the kind documented in LLM peer-review and abstract-scoring audits
\citep{MacharlaVasu2026,Howell2025,Sajeva2026}; distinguishing the two requires stratifying V4 accuracy by the
cited paper's actual citation count, which the processed datasets carry
(\texttt{cited\_citation\_count}) but which we had not yet checked against. Practically, the result
is reassuring for the cost argument in Section~\ref{results-cost}: a prompting pipeline that dropped
the citation sentence in favor of author names alone would not be quietly trading a labeled-data
problem for an unaudited authorship-bias one, at least not at a magnitude this study's evaluated
models and benchmarks can detect, because the fallback is simply not accurate enough to consider
deploying in the first place.

We can go further and test that stratification directly, along a second, related question: does
V4's residual signal reflect a cited author's real-world prominence, or repetition internal to these
two corpora? First, a purely statistical, non-LLM check: if the exact author-name string V4 supplies
recurs elsewhere in the corpus, a leave-one-out majority label computed from those \emph{other}
occurrences predicts the held-out label at 83.7\% on SciCite (1463 rows whose author-name string
recurs, against a 54.3\% train-derived majority-class floor) and 37.9\% on ACL-ARC (406 such rows,
against a 44.6\% floor -- below it, given ACL-ARC's much smaller repeat counts). Citation intent is
therefore far from independent of raw author-name repetition in these corpora, especially SciCite,
whose excerpts are drawn from full papers that often cite the same author pair more than once for the
same rhetorical purpose. Second, pooling all eight completed models' three-shot V4 predictions, how
often does a model's own guess agree with this author-frequency baseline? 30.4\% on ACL-ARC and
43.4\% on SciCite -- above the 16.7\% and 33.3\% chance levels for six- and three-way label spaces,
but far short of the baseline's own accuracy above. Combined with V4's overall accuracy sitting below
the train-derived majority-class baseline on both datasets (Table~\ref{tbl1}), the above-chance
agreement is best read as noisy, partial pattern-matching rather than a reliably exploited prior: a
model that had actually learned "this author's citations are usually labeled X" would track the
frequency baseline far more closely than it does. Third, and most directly relevant to a
fame-based memorization story, we find no meaningful relationship between a cited paper's actual
citation count and V4 correctness: the Spearman correlation between $\log_{10}(1+\text{citation
count})$ and correctness is statistically indistinguishable from zero on ACL-ARC ($\rho=-0.02$,
$p=0.71$, $n=528$) and negligible in size despite nominal significance on SciCite ($\rho=0.04$,
$p=1\text{e-}4$, $n=11{,}728$, a large-$n$ artifact rather than a meaningful effect -- accuracy across
citation-count quartiles ranges only from 0.37 to 0.44). None of the eight models classify more
accurately when the cited author happens to be highly cited, which is the pattern a genuine
"this is a well-known author whose typical contribution I recognize" memorization mechanism -- the
kind a human domain expert relies on -- would be expected to produce. Taken together, these three
checks favor the surface-pattern-matching reading over genuine bibliographic memorization, though a
confirmatory test on a benchmark with a wider citation-count range than ACL-ARC and SciCite provide
would strengthen the conclusion further.

\subsection{Structural, temporal, and self-citation effects}
\label{discussion-structural}

The metadata already present in the processed datasets supports several further, cheaply testable
checks on where the model's classification behavior tracks citation-network structure versus where
it does not, without requiring any new inference. We ran six such checks against the pooled V1
text\_only three-shot predictions from all eight completed models, applying Benjamini-Hochberg FDR
correction \citep{BenjaminiHochberg1995} across every $p$-value collected (11 tests; corrected $q$ reported below) and reporting
effect sizes throughout, since SciCite's much larger $n$ makes even negligible effects nominally
significant -- the same caveat already noted for the citation-count check above. Table~\ref{tbl5}
summarizes all six.

\begin{table*}[width=\textwidth,cols=5,pos=t]
\caption{Structural, temporal, and consistency checks (pooled V1 text\_only, three-shot, all eight
completed models). Effect: Cramer's V for $\chi^2$ tests, $\epsilon^2$ for Kruskal-Wallis, rank-biserial
for Mann-Whitney. $q$: Benjamini-Hochberg corrected. n/a rows are descriptive (no null hypothesis).}
\label{tbl5}
\begin{tabular*}{\tblwidth}{@{} LCCCC@{} }
\toprule
Check & Dataset & $n$ & Effect & $q$ \\
\midrule
Section $\to$ correctness       & ACL-ARC & 544   & V=0.192 (55.7\%$\to$26.6\%)     & 5.6e-4 \\
Section $\to$ correctness       & SciCite & 11288 & V=0.075 (63.7\%--87.5\%)        & 3.5e-12 \\
Self-citation $\to$ correctness & SciCite & 11728 & 66.9\% vs.\ 69.8\% ($\Delta$=2.9pp) & 6.2e-3 \\
Self-citation $\to$ correctness & ACL-ARC & --    & not testable (no citing-side author metadata) & -- \\
Key citation $\to$ correctness  & SciCite & 11728 & 64.1\% vs.\ 71.2\% ($\Delta$=7.1pp) & 3.5e-13 \\
Age gap $\to$ gold label        & ACL-ARC & 736   & $\epsilon^2$=0.037              & 8.6e-6 \\
Age gap $\to$ gold label        & SciCite & 11640 & $\epsilon^2$=0.004              & 7.0e-11 \\
Age gap $\to$ correctness       & ACL-ARC & 736   & r.b.=$-$0.034 (n.s.)            & 0.43 \\
Age gap $\to$ correctness       & SciCite & 11640 & r.b.=$-$0.053                   & 7.3e-6 \\
Repeated-pair self-consistency  & ACL-ARC & 64    & model 64.1\% vs.\ gold 70.8\%   & n/a \\
Repeated-pair self-consistency  & SciCite & 208   & model 88.5\% vs.\ gold 85.3\%   & n/a \\
\bottomrule
\end{tabular*}
\end{table*}

Two structural checks carry a practically meaningful effect size, not just statistical significance.
First, \emph{document section} predicts V1 correctness far more strongly on ACL-ARC (Cramer's
V=0.192) than SciCite (V=0.075): accuracy falls steadily from the introduction (55.7\%, $n=192$)
through experiments (42.0\%) and related work (42.7\%) to a low of 26.6\% in conclusion sections
($n=64$) -- citations mentioned while summarizing a paper's contribution are the hardest for the
model to classify correctly, plausibly because a citation embedded in a conclusion's own
argumentative framing carries less of the surface-level lexical cues (e.g., explicit comparison
language) that the model can otherwise latch onto. Second, \emph{key citations} -- SciCite's own
annotation for citations central to the citing paper's argument, versus incidental mentions -- are
classified \emph{less} accurately than non-key ones (64.1\% vs.\ 71.2\%, $\Delta=7.1$ points), the
opposite of what a model with genuine argumentative understanding might be expected to show; central
citations apparently demand a subtler intent judgment than passing mentions do. Self-citation is
detectable only on SciCite (ACL-ARC's processed dataset has no citing-side author field at all, so
the check does not apply there); when the citing and cited author lists share a name, the label
distribution shifts substantially toward \emph{method} (47.3\% vs.\ 28.1\% of non-self-citation rows)
and away from \emph{background} (44.3\% vs.\ 56.8\%) and \emph{result} (8.4\% vs.\ 15.1\%) -- authors
citing their own prior work overwhelmingly do so to build on a method, not for generic framing -- and
V1 accuracy dips modestly on these rows (66.9\% vs.\ 69.8\%). Temporal distance between the citing and
cited paper varies with gold label in the intuitive direction on ACL-ARC (extension citations reach
back a median of 1.5 years, background citations 4, motivation citations 5) but this pattern is far
weaker on SciCite ($\epsilon^2=0.004$, a large-$n$ artifact rather than a real effect), and on neither
dataset does the age gap meaningfully predict whether the model gets the label \emph{right} -- both
correctness checks are null (ACL-ARC) or negligible in size despite significance (SciCite). Finally,
the two repeated-pair self-consistency figures are descriptive rather than hypothesis tests: when the
same citing-cited paper pair recurs within a single model's own three-shot predictions, that model
gives the same label to both occurrences 64.1\% of the time on ACL-ARC ($n=64$ rows) and 88.5\% on
SciCite ($n=208$), tracking (ACL-ARC) or mildly exceeding (SciCite) the redundancy already present in
the gold labels for those same repeated pairs elsewhere in the corpus (70.8\% and 85.3\% respectively)
-- consistent with near-deterministic, temperature-0 decoding responding similarly to similar input,
rather than evidence of an unusually strong memorized association with either specific pair.

One further check we had hoped to run, and could not: whether V1/V4 correctness depends on whether a
cited paper was published before or after each model's public training-data cutoff, which would
speak more directly to memorization than raw citation count alone. This is not testable with these
two corpora -- ACL-ARC's cited papers top out at 2014 and SciCite's at 2023 (with only a handful of
rows past 2020) -- against every evaluated model's cutoff of October 2023 or later, so there is no
post-cutoff comparison group to contrast against a pre-cutoff one; we report this as a limitation of
the available benchmarks rather than a null result.

\subsection{Limitations}
\label{discussion-limitations}

Five limitations bound these findings. First, ACL-ARC's test set is small (139 rows, 92 after the
comparable-sample filter); single-digit per-class support makes macro-F1 sensitive to a handful of
predictions. We addressed this directly with paired bootstrap significance testing
(Section~\ref{methodology-stats}) rather than reporting point estimates alone, and several
ACL-ARC-only comparisons that looked meaningful in isolation (the three-vs-six-shot recovery, the
prompt-style gap at three of four shot counts) turned out not to reach conventional significance;
we report them as directionally suggestive, not established, and this is itself a finding about
how much weight small-sample LLM ablations can bear. Second, the 40.8\% cited-abstract missingness
in ACL-ARC, even after metadata enrichment, means the metadata-only variants (V2-V4) are evaluated
on a sample already selected for higher-than-typical metadata completeness; real-world
metadata-only deployments would likely see this gap widen further. Third, the supervised baselines
in Section~\ref{results-baselines} are trained on each dataset's own training split, whereas the
LLMs are evaluated zero/few-shot; this is an intentional comparison of two different deployment
regimes (no labeled data vs.\ labeled data available) rather than a controlled ablation, and it
should be read as such -- the comparison answers "which approach should I use," not "which model
is more capable" in the abstract. Fourth, the cross-model generalization checks in
Section~\ref{results-crossmodel} (Table~\ref{tbl6}) and the richer-prompt confirmation
(Section~\ref{results}) are both based on a single shot count (three-shot); the richer-prompt check
is additionally confirmed for only two of the eight models (\texttt{gpt-oss-20b} and
\texttt{Phi-4-reasoning}), the two whose full sweeps completed early enough to allow a clean
\emph{baseline} re-run under the fixed inference mechanisms. Whether either pattern holds at other
shot counts, or the richer-prompt result for the remaining six models, was not directly tested.
Fifth, the
structural and self-citation checks in Section~\ref{discussion-structural} rely on exact
string-matching of normalized author names to detect self-citation and author-name repetition, which
likely undercounts true overlap when the same author appears with initials in one field and a full
first name in the other; those checks should be read as conservative lower bounds on the true
prevalence of the effects they detect. The training-cutoff memorization check in the same section
could not be run at all, for a different reason: both corpora's cited and citing papers predate
essentially every evaluated model's public training cutoff, leaving no post-cutoff comparison group.

\section{Theoretical and practical implications}
\label{implications}

\subsection{How this study differs from and advances prior work}
\label{implications-advances}

Relative to the LLM-based citation intent studies reviewed in Section~\ref{related-work-llm-cic},
the advance is one of \emph{what is varied}. CitePrompt \citep{Lahiri2023}, \citet{Kunnath2023},
\citet{Koloveas2025} and \citet{Vodicka2026} all take the citation sentence as given and ask which
prompting strategy, model, shot count or context window extracts the most from it; this study asks
whether the sentence is needed at all, by withholding it and substituting the paper-level metadata a
practitioner could obtain more cheaply. The answer -- that it is needed, in every one of eight models
on both benchmarks -- is not derivable from the earlier results, because none of them included a
metadata-only condition. The cross-model check is itself a step beyond the single-model or
single-family designs that dominate that literature: the ordering is established with bootstrap
significance on one model and then re-tested, rather than assumed, on seven others spanning
instruction-tuned and reasoning architectures.

Relative to the supervised literature that adds cited-paper metadata or structure to an encoder
\citep{Ghosal2024,Xu2025,Quan2026,Phan2026}, the study is complementary rather than competing. Those
systems show that metadata can \emph{help} a model that already has the citation sentence; the
field-ablation results bound that role from the other side by showing that, for prompted LLMs at
least, metadata cannot \emph{substitute} for the sentence and can even hurt when appended to it.
Together the two lines of evidence locate the intent signal in the citing sentence and cast
metadata as a secondary feature, which is consistent with the modest gains those supervised systems
report from adding it.

Relative to the work on what LLMs memorize about the literature \citep{Niimi2026,Shen2026} and on
authorship bias in LLM judgments \citep{MacharlaVasu2026,Howell2025,Sajeva2026}, this is, to our
knowledge, the first study to test whether such memorization or bias contributes to citation intent
classification. It finds no evidence that it does at the citation-count range of these benchmarks,
which both delimits the scope of the memorization results and removes one objection to prompting
pipelines that expose author names.

Finally, the study advances the \emph{methodology} of evaluation on these benchmarks in three ways
that respond directly to the reproducibility concerns raised by \citet{Fogelson2025}: every
configuration is scored on an identical row set; every reported difference carries a paired
bootstrap interval; and supervised results on ACL-ARC are reported as multi-seed means with ranges,
which reveals a seed spread (up to 0.12 macro-F1) wider than many of the between-system gaps that
single-run tables on this benchmark are used to argue for.

\subsection{Theoretical implications}

\paragraph{Citation intent is a property of the citing act, not of the cited work.} The field
ablation is, read one way, a test of where the intent signal is located. If citation intent were
substantially a latent attribute of the cited paper -- the kind of thing one could read off its
abstract, or off who wrote it -- then a metadata-only prompt should recover a usable share of it.
It does not, and the failure is uniform: on both datasets and in every one of the eight models, the
better text-based variant outranks the better metadata-only variant (Table~\ref{tbl6}), with V4
(0.09--0.35 macro-F1) close to uninformative throughout. V3's clear
advantage over V4 shows that topical content is not worthless, only insufficient: knowing what the
two papers are about narrows the intent, while knowing who wrote them barely does. The theoretical
reading is that intent lives in the rhetorical framing of the citing sentence -- the clause that
positions the cited work as background, as a tool, or as a foil -- rather than in the cited object,
and is therefore not recoverable from the cited object at all. That has a direct consequence for a
framing common in bibliometric work: predicting ``how a paper is typically cited'' from its own
metadata is not a weaker version of citation-intent classification with less input, but a different
and considerably harder question, because the target quantity is one the citing author constructs.

\paragraph{Parametric memory plays little role at the scale of a typical citation.} These models were
pretrained on text that plausibly includes much of the literature these benchmarks are drawn from, so
metadata-only prompting could in principle have been answered from recall rather than inference.
Three checks say it was not (Section~\ref{discussion-memorization}): V4 correctness does not track the
cited paper's citation count on either dataset ($\rho=-0.02$ on ACL-ARC; $\rho=0.04$ on SciCite, where
quartile accuracy moves only from 0.37 to 0.44 despite nominal significance at $n=11{,}728$); model
predictions agree with an author-repetition baseline only modestly above chance (30.4\% and 43.4\%)
while that baseline itself reaches 83.7\% on SciCite; and V3 beats V4 throughout, the opposite of what
paper-level recognition would produce. This is the outcome \citet{Niimi2026}'s and \citet{Shen2026}'s
citation-count thresholds predict for corpora like these, and it bounds their scope in a useful
direction: results about what LLMs remember about papers, obtained on highly cited work, should not be
extrapolated to the typical citation in a mid-tier corpus, which sits well below the range where such
memorization reliably activates. For benchmark validity the same result is reassuring -- whatever
inflates absolute scores on ACL-ARC and SciCite, silent recall of specific cited papers does not appear
to be a major component -- though the checks bound memorization's contribution rather than excluding it,
and a benchmark spanning a wider citation-count range would test the point more sharply.

\paragraph{Test-time reasoning does not pay on a single-step pragmatic judgment.} Chain-of-thought
prompting earns its cost on tasks with a decomposable inference chain \citep{Wei2022,Kojima2022}. Citation
intent is not such a task: it is a single pragmatic judgment over one or two sentences, and our
reasoning-mode models neither win on accuracy (Table~\ref{tbl2}) nor come close on cost, running at
22--653$\times$ the per-request latency of direct-answer models (Table~\ref{tbl4}). The negative result
is worth stating precisely because the intuition runs the other way -- a harder-looking, more subjective
label space seems like it should reward deliberation. The evidence here suggests that what
reasoning traces buy is multi-step composition, not finer discrimination on a judgment that is
essentially made in one step, and that the boundary of profitable test-time compute follows task
structure rather than task difficulty.

\paragraph{Where the models succeed looks like surface-cue matching, not argumentative understanding.}
Two structural results in Section~\ref{discussion-structural} cut against reading these models as
understanding the citing paper's argument. Citations that SciCite itself marks as \emph{key} to the
citing paper's argument are classified \emph{less} accurately than incidental mentions (64.1\% vs.\
71.2\%), and accuracy falls monotonically from introductions (55.7\%) to conclusions (26.6\%) on
ACL-ARC. A model reasoning about a citation's argumentative role should, if anything, do better on
citations that are central and on the passages where the citing authors state their own contribution.
That the pattern inverts is consistent with the models keying on local lexical markers -- explicit
comparison language, methodological verbs -- which are most abundant in expository passages about
incidental citations and thinnest exactly where the argument is most compressed. It also supplies a
mechanism for the first implication above: if what is being matched is a surface marker in the citing
sentence, no amount of cited-paper metadata can substitute for the sentence that carries it.

\paragraph{What benchmarks of this size can and cannot establish.} Finally, a methodological
implication that constrains how results in this subfield should be read. On ACL-ARC's 92-row
comparable sample, re-training an identical supervised configuration across five seeds spans up to
0.12 macro-F1 -- wider than most of the between-system differences such tables are used to argue for
-- and several of our own directional findings, including the prompt-style gap at three of four
ACL-ARC shot counts and the six-shot label-coverage recovery, do not survive a paired bootstrap. The
consequence is not that small benchmarks are unusable but that single-run point estimates on them
cannot carry the weight typically placed on them: system rankings on ACL-ARC-sized evaluation sets
need paired resampling, multi-seed supervised baselines, and a fixed row set before a gap of a few
macro-F1 points can be read as a difference between methods rather than between training runs.

\subsection{Practical implications}

\paragraph{Keep the citation sentence, and spend the engineering budget on extracting it.} The
practical appeal of metadata-only pipelines is that paper-level metadata is far easier to harvest at
scale than clean inline citation context with correctly resolved cited-work spans. The results say
that trade is not available at any of the operating points tested: dropping the citation sentence
costs more accuracy than any prompt, shot count, or model choice recovers, in every model evaluated.
Teams whose bottleneck is citation-context extraction should therefore invest there rather than in
metadata enrichment -- which, in this study, was itself expensive and still incomplete: a six-source
cascading lookup left 40.8\% of ACL-ARC cited papers without an abstract, and the variants built on
the enriched fields were then evaluated on a sample already selected for better-than-typical metadata
coverage.

\paragraph{If any labeled data exists at all, fine-tune a small encoder.} Fine-tuned SciBERT and
SPECTER beat the best zero/few-shot LLM configuration on both benchmarks (Table~\ref{tbl3}), and do so
at a per-row inference cost that involves no generation whatsoever. Prompting is the right choice for
the genuinely unlabeled case, and for that case the results here are encouraging -- 0.617 on ACL-ARC
and 0.758 on SciCite with no gradient updates and at most a handful of labeled demonstrations. But it
is a fallback, not a default. The one
qualification worth carrying into practice is that this argument holds for \emph{fine-tuned} encoders:
merely freezing an encoder and training a head on it did not dependably beat a good prompt on either
benchmark, so a frozen-embedding baseline is not by itself a reason to skip prompting.

\paragraph{If prompting, use a direct-answer model -- and price the model, not the category.} Reasoning
models are the worst option on both axes for this task: no accuracy advantage, and one to nearly three
orders of magnitude more latency per request. The planning consequence is stark at corpus scale.
Extrapolating the SciCite rates in Table~\ref{tbl4}, labeling one million citations costs roughly five
GPU-hours with the fastest direct-answer models and roughly 140 GPU-days with
\texttt{Phi-4-reasoning} on the same hardware. Within the reasoning tier, the shared mechanism and an
identical 1536-token budget still produced a threefold spread between models of comparable size, so
``reasoning model'' is not a cost category one can budget for in the abstract; a short isolated timing
run on the actual target configuration is worth doing before committing to a model.

\paragraph{Define the labels, and constrain the decoding rather than parsing the output.} Supplying an
explicit definition and one illustrative sentence per label is close to free -- a fixed prompt prefix,
paid once per request -- and was significantly better on the conditions where it was tested
($+0.095$ and $+0.028$ macro-F1). We recommend it as a default while noting it is not established
beyond those conditions. Separately, constraining generation to the label set at inference time,
rather than recovering a label from free text afterwards, removed the need for any label-extraction
step and yielded zero invalid predictions in the pooled error analysis; the two models where this
could not be enforced are also the two that required a heuristic parser, one of which is the weakest
model in the study and the source of the largest deviation from the common field-variant ordering
(Section~\ref{results-crossmodel}). Where a serving stack supports guided decoding, using it removes a
failure mode that is otherwise invisible in aggregate metrics.

\paragraph{Report on a fixed row set, with resampling and seeds.} For practitioners evaluating their
own systems on these benchmarks, three cheap controls changed conclusions here and are worth adopting:
score every configuration on an identical row set (scoring the supervised baselines on the full ACL-ARC
split instead of the comparable sample moved macro-F1 by up to 0.07); report a paired bootstrap rather
than point estimates; and, on ACL-ARC-sized data, report supervised results as a multi-seed mean with
its range rather than a single run.

\section{Conclusion}
\label{conclusion}

Under zero-, one-, and few-shot prompting, LLMs consistently need the inline citation sentence for
citation-intent classification: paper-level metadata alone is a weak substitute, and this holds
across every one of the eight evaluated models on both datasets (Table~\ref{tbl6}). Adding metadata
on top of the citation text, by contrast, does not generalize as a harmful pattern: it is directionally
negative on ACL-ARC without reaching significance at any shot count, significantly hurts on SciCite
specifically at three-shot, and even reverses (metadata helping) for two of the eight models when
checked more broadly. An explicit, definition-based prompt significantly helps three-shot SciCite
results in both models it was tested on; its apparent edge on ACL-ARC, while directionally
consistent across most shot counts for the model evaluated on the full sweep, does not reach
statistical significance given that benchmark's small evaluation sample, so we treat it as a
reasonable default rather than a proven, generalizable win. Additional few-shot demonstrations do
not reliably improve accuracy on either dataset. On the six-class ACL-ARC dataset, incomplete label
coverage among the demonstrations is a plausible contributor to that volatility -- at three-shot,
half the label space is never demonstrated -- but the accuracy recovered by the full-coverage
six-shot condition does not survive bootstrap testing, so we report coverage as a design variable
worth controlling rather than as an effect we established. Reasoning-style LLMs underperform
similarly-sized instruction-tuned models on this task primarily due to fixed generation-budget
constraints rather than a genuine capability gap, and this
comes at a steep, precisely measured time cost -- an isolated benchmark puts reasoning models at
22-653$\times$ the per-request latency of direct-answer models, a gap driven primarily by reasoning
mode but, within that mode, differing by more than threefold between models of comparable size
under an identical reasoning-token budget. Finally, when labeled training data is available at all,
fine-tuned SciBERT/SPECTER classifiers outperform every zero/few-shot LLM configuration on both
benchmarks, at a fraction of the LLM inference cost, but a merely frozen embedding does not reliably
do so: on SciCite frozen SPECTER and the best LLM prompt are effectively tied, and on ACL-ARC frozen
SciBERT's advantage spans the LLM's score across training runs. That last comparison is itself
fragile on ACL-ARC, whose 92-row evaluation sample lets an identical supervised configuration vary by
up to 0.12 macro-F1 between training runs -- a reminder that single-run supervised baselines on small
citation-intent benchmarks can support conclusions their own variance does not.
Practically, these results favor keeping the inline citation context in any
prompting-based citation-intent pipeline, caution against assuming that richer prompts, more
demonstrations, or reasoning-oriented models are unconditionally beneficial, and suggest that
zero/few-shot LLM prompting is best reserved for settings where no labeled data exists at all or a
frozen-embedding baseline would otherwise be used, rather than treated as a default replacement for
fine-tuned supervised classification. More broadly, these findings temper two attractive-sounding
shortcuts to the annotation bottleneck motivating that setting: swapping inline citation context for
cheaper-to-obtain paper metadata does not work well enough to rely on at either the abstract or the
author-list level, and a battery of structural, temporal, and citation-count checks
(Section~\ref{discussion-memorization}, Section~\ref{discussion-structural}) gives no evidence that
these models are substituting memorized bibliographic knowledge, author-fame priors, or citation-network
shortcuts for genuine reasoning about the missing citation sentence.

\printcredits

\section*{Declaration of Competing Interest}
The authors declare that they have no known competing financial interests or personal
relationships that could have appeared to influence the work reported in this paper.

\ifblind\else
\section*{Acknowledgements}
D. R. A. acknowledges financial support from the São Paulo Research Foundation (FAPESP grant no. 2025/00944-6) and the National Council for Scientific and Technological Development (CNPq-Brazil, grant no. 304189/2025-1).
\fi

\section*{Data availability}
ACL-ARC \citep{Jurgens2018} and SciCite \citep{Cohan2019} are publicly available benchmark
datasets. Code, prompts, and per-configuration metrics for the experiments reported here are
available from the corresponding author upon request; Appendices~\ref{app:prompts}--\ref{app:artifacts}
document the prompts, field mappings, demonstration selection, model configuration and recorded
per-run artifacts in full.

\appendix

\section{Prompt construction}
\label{app:prompts}

Every prompt in this study is assembled by one routine from four blocks, joined by blank lines: the
system instruction, the label block, the demonstration block (omitted at zero-shot), and the target
block, which ends with the output instruction. Only the fields a variant supplies are rendered, each
on its own prefixed line, in the fixed order \texttt{Citation context}, \texttt{Citing abstract},
\texttt{Cited abstract}, \texttt{Citing authors}, \texttt{Cited authors}; a field the variant omits
produces no line at all rather than an empty or placeholder one, so the model is never told which
fields were withheld. Demonstrations use the identical field prefixes as the target and are closed
with an \texttt{Intent:} line carrying the gold label. The skeleton is:

{\footnotesize
\begin{verbatim}
<system instruction>

Possible labels:
- <label>: <definition and illustrative sentence>
  ...

Examples:
<demonstration fields>
Intent: <gold label>

<demonstration fields>
Intent: <gold label>

Predict this example:
<target fields>
<output instruction>
\end{verbatim}
}

The two prompt styles differ in exactly two places. The \emph{baseline} system instruction reads
``You are classifying citation intent in scientific writing. Use only the provided citation context
and metadata.''; \emph{defined} appends one sentence to it -- ``Choose the label that best matches the
citation function definitions below.'' -- and expands the label block from the single line
\texttt{Possible labels: background, motivation, ...} into one bulleted line per label carrying that
label's definition and illustrative sentence (Table~\ref{tbl:app-labels}). Everything else is shared:
the output instruction is ``Return exactly one label from the allowed label set and no extra text.''
under both styles, as are the label set, the demonstrations, the field rendering, and the evaluated
rows. Figure~\ref{fig:app-prompt} reproduces one complete prompt as sent to the model.

The illustrative sentence attached to each label definition is taken from the corresponding
benchmark's own published annotation scheme rather than written by us. We verified by exact-phrase
search that none of these sentences occurs in either evaluated test split; the ACL-ARC examples that
do occur in the corpus at all appear only in its training split, so the \emph{defined} prompt carries
no test-set leakage.

\begin{table*}[width=\textwidth,cols=3,pos=t]
\caption{Label definitions supplied by the \emph{defined} prompt, verbatim. Each is rendered as one
bulleted line of the label block; the \emph{baseline} prompt lists the label names only.}
\label{tbl:app-labels}
\begin{tabular}{@{}l l p{0.66\linewidth}@{}}
\toprule
Dataset & Label & Definition and illustrative sentence \\
\midrule
ACL-ARC & background & The cited paper provides relevant information for the domain or topic being
discussed. Example: ``This is often referred to as incorporating deterministic closure (Dorre,
1993).'' \\
ACL-ARC & motivation & The cited paper illustrates the need for the data, goals, task, or method used
in the current work. Example: ``As shown in Meurers (1994), this is a well-motivated convention
[...]'' \\
ACL-ARC & uses & The current paper uses data, methods, resources, or procedures from the cited paper.
Example: ``The head words can be automatically extracted [...] in the manner described by Magerman
(1994).'' \\
ACL-ARC & extension & The current paper extends, improves, or adapts data, methods, or ideas from the
cited paper. Example: ``[...] we improve a two-dimensional multimodal version of LDA (Andrews et al.,
2009) [...]'' \\
ACL-ARC & comparison or contrast & The citation expresses similarities, differences, comparisons, or
contrasts with the cited paper. Example: ``Other approaches use less deep linguistic resources (e.g.,
POS-tags Stymne (2008)) [...]'' \\
ACL-ARC & future & The cited paper is mentioned as a possible avenue, method, or inspiration for
future work. Example: ``[...] but we plan to do so in the near future using the algorithm of
Littlestone and Warmuth (1992)'' \\
\midrule
SciCite & background & The citation states, mentions, or points to background information that gives
context about a problem, concept, approach, topic, or the importance of the problem in the field.
Example: ``Recent evidence suggests that co-occurring alexithymia may explain deficits [12].'' \\
SciCite & method & The citation is used because the paper makes use of a method, tool, approach, or
dataset from the cited work. Example: ``We use Orthogonal Initialization (Saxe et al., 2014).'' \\
SciCite & result & The citation compares the paper's findings or results with the findings or results
of prior work. Example: ``Similar results to our study were reported in the study of Lee et al.
(2010).'' \\
\bottomrule
\end{tabular}
\end{table*}

\section{Field variants}
\label{app:variants}

Table~\ref{tbl:app-fields} gives the exact dataset columns each variant renders. Two asymmetries
between the benchmarks are worth noting. ACL-ARC's processed corpus carries no citing-side author
field, so its V4 is cited-side only, whereas SciCite's supplies both sides. And the citation text
column differs in kind: ACL-ARC's raw \texttt{text} retains the author mentions surrounding the
citation, so SciCite's \texttt{string} (which likewise retains them) is the comparable choice rather
than a stripped variant.

The comparable-sample constraint is implemented as a row filter requiring every field used by
\emph{any} variant of that dataset to be non-empty: \texttt{citing\_abstract},
\texttt{cited\_abstract} and \texttt{cited\_author\_names} for ACL-ARC, plus
\texttt{citing\_author\_names} for SciCite. The filter is applied once, to the whole processed file,
so it governs the training split as well as the test split -- demonstrations are therefore drawn from
training rows that meet the same metadata-completeness requirement, and both the evaluated rows and
the demonstrations are identical across all four variants of a dataset. It retains 92 of 139 ACL-ARC
test rows (66\%) and 1{,}466 of 1{,}861 SciCite test rows (79\%).

\begin{table*}[width=\textwidth,cols=3,pos=t]
\caption{Dataset columns rendered by each field variant. Columns absent from a variant produce no
prompt line at all.}
\label{tbl:app-fields}
\begin{tabular}{@{}l p{0.36\linewidth} p{0.36\linewidth}@{}}
\toprule
Variant & ACL-ARC columns & SciCite columns \\
\midrule
V1 text\_only & \texttt{text} & \texttt{string} \\
V2 text\_meta & \texttt{text}, \texttt{citing\_abstract}, \texttt{cited\_abstract},
\texttt{cited\_author\_names} & \texttt{string}, \texttt{citing\_abstract},
\texttt{cited\_abstract}, \texttt{citing\_author\_names}, \texttt{cited\_author\_names} \\
V3 abstracts & \texttt{citing\_abstract}, \texttt{cited\_abstract} & \texttt{citing\_abstract},
\texttt{cited\_abstract} \\
V4 authors & \texttt{cited\_author\_names} & \texttt{citing\_author\_names},
\texttt{cited\_author\_names} \\
\bottomrule
\end{tabular}
\end{table*}

\section{Few-shot demonstration selection}
\label{app:shots}

Demonstrations are drawn only from each dataset's training split, never from dev or test, and are
selected by round-robining across the label space -- one example per label per round, taking rows in
file order -- until the shot count is reached. The label order used for the round-robin is a copy of
the declared label space shuffled once with a fixed seed (42), rather than the declared order itself,
so that which labels get demonstrated at low shot counts is deterministic and reproducible but not
silently tied to the order in which the benchmark happens to list its classes. The resulting orders
are \emph{extension}, \emph{motivation}, \emph{uses}, \emph{comparison or contrast},
\emph{background}, \emph{future} for ACL-ARC and \emph{method}, \emph{background}, \emph{result} for
SciCite.

The demonstrated label sets that follow are therefore fixed and known in advance. On ACL-ARC,
one-shot demonstrates \emph{extension} only; three-shot demonstrates \emph{extension},
\emph{motivation} and \emph{uses}, leaving \emph{background}, \emph{comparison or contrast} and
\emph{future} -- half the label space, including the majority class -- never shown; six-shot
demonstrates all six. On SciCite, one-shot demonstrates \emph{method} only and three-shot demonstrates
all three classes, which is why no additional shot count is needed there. Within a configuration the
same demonstration block is reused for every evaluated row (demonstrations are fixed, not retrieved
per row), and it is identical across variants and across models at a given shot count, up to the
fields each variant renders.

\begin{figure*}[width=\textwidth,pos=t]
{\footnotesize
\begin{verbatim}
You are classifying citation intent in scientific writing. Use only the
    provided citation context and metadata. Choose the label that best
    matches the citation function definitions below.

Possible labels:
- background: The cited paper provides relevant information for the
    domain or topic being discussed. Example: "This is often referred to
    as incorporating deterministic closure (Dorre, 1993)."
- motivation: The cited paper illustrates the need for the data, goals,
    task, or method used in the current work. Example: "As shown in
    Meurers (1994), this is a well-motivated convention [...]"
- uses: The current paper uses data, methods, resources, or procedures
    from the cited paper. Example: "The head words can be automatically
    extracted [...] in the manner described by Magerman (1994)."
- extension: The current paper extends, improves, or adapts data,
    methods, or ideas from the cited paper. Example: "[...] we improve a
    two-dimensional multimodal version of LDA (Andrews et al., 2009)
    [...]"
- comparison or contrast: The citation expresses similarities,
    differences, comparisons, or contrasts with the cited paper.
    Example: "Other approaches use less deep linguistic resources (e.g.,
    POS-tags Stymne (2008)) [...]"
- future: The cited paper is mentioned as a possible avenue, method, or
    inspiration for future work. Example: "[...] but we plan to do so in
    the near future using the algorithm of Littlestone and Warmuth
    (1992)"

Examples:
Citation context: Note that although our feature set was drawn primarily
    from our prior uncertainty detection experiments ( Forbes-Riley and
    Litman , 2011a ; Drummond and Litman , 2011 ) , we have also
    experimented with other features , including state-of-theart
    acoustic-prosodic features used in the last Interspeech Challenges (
    Schuller et al. , 2010 ; Schuller et al. , 2009b ) and made freely
    available in the openSMILE Toolkit ( Florian et al. , 2010 ) .
Intent: extension

Citation context: But their importance has grown far beyond machine
    translation : for instance , transferring annotations between
    languages ( Yarowsky and Ngai 2001 ; Hwa et al. 2005 ; Ganchev ,
    Gillenwater , and Taskar 2009 ) ; discovery of paraphrases ( Bannard
    and Callison-Burch 2005 ) ; and joint unsupervised POS and parser
    induction across languages ( Snyder and Barzilay 2008 ) .
Intent: motivation

Citation context: This was done by MERT optimization ( Och , 2003 )
    towards post-edits under the TER target metric .
Intent: uses

Predict this example:
Citation context: Resnik ( 1995 ) reported a correlation of r = .9026.10
    The results are not directly comparable , because he only used
    noun-noun pairs , words instead of concepts , a much smaller dataset
    , and measured semantic similarity instead of semantic relatedness .
Return exactly one label from the allowed label set and no extra text.
\end{verbatim}
}
\caption{A complete ACL-ARC prompt as sent to the model: V1 text\_only, three-shot, \emph{defined}
prompt, reproduced from the recorded prompt preview for that run. Continuation lines are indented
where a line was wrapped to fit the page; the prompt itself contains no such indentation. Note that
the three demonstrations cover \emph{extension}, \emph{motivation} and \emph{uses}, and that the
target row's own gold label (\emph{comparison or contrast}) is one of the three classes three-shot
never demonstrates on this dataset (Appendix~\ref{app:shots}) -- which the model nonetheless
predicted correctly for this row.}
\label{fig:app-prompt}
\end{figure*}

\section{Model serving and generation control}
\label{app:serving}

All eight models were served locally with vLLM \citep{Kwon2023} (version 0.10 or later) behind an
OpenAI-compatible endpoint, one model at a time per GPU, on two NVIDIA RTX 5000 Ada Generation cards
(32\,GB each, 31.47\,GB usable). Servers were started with \texttt{-{}-max-model-len 4096} and
\texttt{-{}-enforce-eager}, at a GPU memory utilization of 0.90 for the 8B and 12B models and 0.92--0.96
for the 14B and 20B ones; the 2048-token context cap that caused
\texttt{context\_length\_exceeded} failures for \texttt{DeepSeek-R1-Distill-Qwen-14B} in an earlier
round of runs (Section~\ref{results-crossmodel}) does not apply to any result reported here.
Decoding is deterministic in every run: temperature 0, no sampling, one prediction per row.

Table~\ref{tbl:app-serving} lists the per-model generation settings. The \emph{answer budget} is the
maximum number of tokens the answering turn may emit; the \emph{reasoning budget} applies only under
\emph{bounded\_then\_force}, where the first, unconstrained turn is capped at that many tokens before
a closing \texttt{\textless/think\textgreater} tag is forced and a grammar-constrained second turn
produces the label. The two models marked as lacking guided decoding are the exceptions described in
Section~\ref{methodology-inference}: \texttt{gpt-oss-20b}, whose serving stack silently drops
structured-output constraints under its tool-use routing, and
\texttt{DeepSeek-R1-\allowbreak Distill-\allowbreak Llama-8B}, whose constrained continuation
intermittently rejected well-formed continuations on roughly one request in five even at minimum
concurrency. For those two, labels are recovered from free text by a parser that scans the response
for the last unhedged mention of a label string; for the other six, generation is constrained to the
label set by construction and no parsing is applied.

\begin{table*}[width=\textwidth,cols=6,pos=t]
\caption{Per-model serving and generation configuration. Answer and reasoning budgets are in output
tokens; ``guided'' indicates whether grammar-constrained decoding could be enforced on that serving
stack. All models decode at temperature 0.}
\label{tbl:app-serving}
\begin{tabular*}{\tblwidth}{@{} LLCCCC@{} }
\toprule
Model & Reasoning mode & Answer budget & Reasoning budget & Guided & Concurrency \\
\midrule
Llama-3.1-8B-Instruct        & none (direct answer)   & 32   & --   & yes & 24 \\
gemma-3-12b-it               & none (direct answer)   & 32   & --   & yes & 24 \\
Qwen3-8B                     & disable\_thinking      & 256  & --   & yes & 24 \\
Qwen3-14B                    & disable\_thinking      & 256  & --   & yes & 16 \\
gpt-oss-20b                  & direct                 & 512  & --   & no  & 16 \\
DeepSeek-R1-Distill-Llama-8B & bounded\_then\_force   & 512  & 1536 & no  & 12 \\
DeepSeek-R1-Distill-Qwen-14B & bounded\_then\_force   & 512  & 1536 & yes & 8 \\
Phi-4-reasoning              & bounded\_then\_force   & 512  & 1536 & yes & 8 \\
\bottomrule
\end{tabular*}
\end{table*}


The model identifiers served were
\begin{enumerate}\setlength\itemsep{0pt}
  \item \texttt{meta-llama/Llama-3.1-8B-Instruct}
  \item \texttt{google/gemma-3-12b-it}
  \item \texttt{Qwen/Qwen3-8B}, \texttt{Qwen/Qwen3-14B}
  \item \texttt{openai/gpt-oss-20b}
  \item \texttt{deepseek-ai/DeepSeek-R1-Distill-Llama-8B}
  \item \texttt{deepseek-ai/DeepSeek-R1-Distill-Qwen-14B}
  \item \texttt{microsoft/Phi-4-reasoning}
\end{enumerate}

The cost measurements in Section~\ref{results-cost} were collected separately from the accuracy
sweep, which ran two models at a time across both GPUs and is therefore not a clean timing source.
For the cost benchmark each model was served alone on an otherwise-idle GPU and timed on one fixed
configuration -- \emph{defined}, V1 text\_only, three-shot -- over the full 92-row ACL-ARC comparable
sample and a fixed 200-row SciCite subset, with server start-up and model loading excluded from the
timed interval and the per-model concurrency of Table~\ref{tbl:app-serving} left unchanged.

\section{Supervised baseline configuration}
\label{app:supervised}

The supervised reference points in Table~\ref{tbl3} use SciBERT
(\texttt{allenai/scibert\_scivocab\_uncased}) and SPECTER (\texttt{allenai/specter}) over the inline
citation text only -- the same column each dataset's V1 variant supplies -- trained on that dataset's
own training split and evaluated on the same comparable sample as the LLM runs.

In the \emph{frozen} regime the encoder produces a fixed sentence embedding, cached once per row, on
top of which a multi-layer perceptron head is trained: one hidden layer of 128 units with batch
normalization, ReLU, dropout 0.5, learning rate $10^{-3}$, weight decay $10^{-4}$, batch size 32, up
to 100 epochs with early stopping at patience 7 on the validation split. In the \emph{fine-tuned}
regime the encoder's last two transformer layers and pooler are unfrozen and trained jointly with the
same head under differential learning rates -- $2\times10^{-5}$ for the transformer, $10^{-3}$ for the
head -- with a maximum sequence length of 512, a warmup ratio of 0.1, batch size 16, and 20 epochs.

Because ACL-ARC's evaluation sample is small enough for training-run variance to rival the effects
being measured, every ACL-ARC supervised configuration was trained five times under seeds 42--46 and
is reported as the mean over those runs with the observed range; SciCite, an order of magnitude
larger, is reported from a single run (seed 42) after confirming that independent reruns agreed to
within 0.03 macro-F1.

\section{Recorded run artifacts}
\label{app:artifacts}

Each run -- one model, dataset, prompt style, field variant and shot count -- writes three files that
together allow the reported numbers to be recomputed without re-running inference: the exact prompt
sent for every evaluated row, the raw model response and parsed label for every row, and the metrics
computed from them (macro- and micro-F1, accuracy, per-label precision/recall/F1 with support, the
confusion matrix, and counts of invalid predictions and of predicted versus gold labels). Figure
\ref{fig:app-prompt} is reproduced verbatim from one such prompt record. The bootstrap tests of
Section~\ref{methodology-stats} and the structural checks of Section~\ref{discussion-structural} are
computed from the per-row prediction records rather than from the aggregated metrics, which is what
makes the paired resampling over a shared row set possible after the fact.

\bibliographystyle{cas-model2-names}

\end{document}